\documentclass[11pt,letterpaper]{article}

\usepackage[margin=1in]{geometry}
\usepackage{amsmath,amssymb,amsthm,mathtools}
\usepackage{booktabs,tabularx,array}
\usepackage{enumitem}
\usepackage[ruled,linesnumbered,vlined]{algorithm2e}
\usepackage{xcolor}
\usepackage{tikz}
\usetikzlibrary{arrows.meta,calc,positioning,decorations.pathreplacing}
\usepackage{microtype}
\usepackage{placeins}
\usepackage[hidelinks]{hyperref}

\newtheorem{theorem}{Theorem}[section]
\newtheorem{lemma}[theorem]{Lemma}

\newtheorem{corollary}[theorem]{Corollary}
\newtheorem{definition}[theorem]{Definition}

\theoremstyle{definition}

\DontPrintSemicolon
\SetAlgoNlRelativeSize{-1}
\SetAlgoInsideSkip{smallskip}
\SetKwProg{Proc}{Procedure}{}{end}

\newcommand{\eps}{\varepsilon}
\newcommand{\Prb}{\mathop{\mathrm{Pr}}}
\newcommand{\E}{\mathop{\mathbb{E}}}
\newcommand{\N}{\mathop{\mathbb{N}}}
\newcommand{\one}{\mathbf{1}}
\newcommand{\polylog}{\operatorname{polylog}}
\newcommand{\negl}{\operatorname{negl}}
\newcommand{\neigh}{\sim}
\newcommand{\Insert}{\mathsf{Insert}}
\newcommand{\Extract}{\mathsf{ExtractMin}}
\newcommand{\view}{\mathsf{Access}}
\newcommand{\op}{\mathsf{op}}

\newcommand{\seen}{\mathsf{seen}}

\title{Differentially Oblivious Priority Queues with Sublogarithmic Cost}
\author{%
  Zimai Guo\textsuperscript{1},
  Cheng Hong\textsuperscript{2},
  Zhechen Li\textsuperscript{2,}\thanks{Authors are listed in lexicographical order. Zhechen Li is the corresponding author.}\\[0.5em]
  {\small \textsuperscript{1}\textnormal{School of Computer Science, Peking University, Beijing, China}\quad \textsuperscript{2}\textnormal{Ant Group, Beijing, China}}\\
  {\small \texttt{\{endergzm,lizhechen\}@pku.edu.cn}, \texttt{vince.hc@antgroup.com}}
}
\date{}

\begin{document}
\setcounter{footnote}{1} 
\maketitle

\begin{abstract}
In this paper, we study exact online integer priority queues under differential obliviousness in the balls-and-bins model. Given a public upper bound $T$ on the number of operations and assuming that the priorities are bounded by a polynomial in $T$, we construct an $(\eps,\delta)$-differentially oblivious priority queue supporting insertion and minimum extraction, which requires $\polylog(T)$ private memory. For constant $\varepsilon>0$ and $\delta=\exp(-\Theta(\log^2 T))$, our construction incurs $O(\log T/\log\log T)$ expected amortized total work per operation, which breaks the $\Omega(\log T)$ barrier for fully oblivious priority queues~[Jacob et al., SODA'2019]. Further, we prove a matching lower bound for $(\eps,\delta)$-differentially oblivious priority queues with $\varepsilon=O(1)$ and $\delta=o(1)$, which indicates that our construction is asymptotically optimal in terms of total work.

\end{abstract}

\section{Introduction}
\label{sec:introduction}

Priority queues (or heaps) are fundamental data structures in the design of algorithms. They support insertion and extraction of a minimum-priority record and are central to algorithms for shortest paths and minimum spanning trees~\cite{FredmanTarjan1987}. Starting from an empty queue, for an online sequence of at most $T$ operations, an ordinary heap (e.g., binary heap, Fibonacci heap) supports both operations in $O(\log T)$ runtime~\cite{Williams1964}. This logarithmic cost is optimal in the comparison-based model, since insertions followed by minimum extractions sort the inserted records.

In commonly used constructions, the memory access traces on RAM induced by operations depend on the stored priorities. Observing these memory accesses may therefore reveal information about the priorities even when the records themselves are encrypted. Full obliviousness~\cite{GoldreichOstrovsky1996} protects against this leakage by requiring access patterns to be indistinguishable for operation sequences with the same public parameters. Shi's Path Oblivious Heap~\cite{Shi2020PathObliviousHeap} and the bandwidth-optimal heap by Jafargholi, Larsen, and Simkin~\cite{JafargholiLarsenSimkin2021} give elegant oblivious priority queues with $O(\log T)$ amortized word transfers per operation. Their per-operation cost matches the optimal runtime of non-oblivious comparison-based priority queues.

For polynomially bounded integer priorities, non-oblivious priority queues can achieve $O(1)$ amortized operation time. Integer priority queues~\cite{Thorup2000} exploit word operations beyond comparisons. With priorities bounded by $T^{O(1)}$ and $w=\Theta(\log T)$-bit words, radix sort sorts $T$ priorities in $O(T)$ time, bypassing the $\Omega(T\log T)$ barrier for comparison sorting. Thorup's reduction from priority queues to sorting then yields the constant amortized bound~\cite{Thorup2007}. However, this improvement does not extend to full obliviousness. In the standard online model with observable call boundaries, Jacob, Larsen, and Nielsen~\cite{JacobLarsenNielsen2019} prove an $\Omega(\log T)$ expected amortized probe lower bound, even for polynomially bounded integer priorities and polylogarithmic private memory.

This gap motivates us to consider differential obliviousness (DO), which relaxes full obliviousness by requiring $(\eps,\delta)$-differential privacy~\cite{Dwork06D} for the access patterns of neighboring inputs~\cite{ChanCCMS2019,ChanChungMaggsShi2022}. Chan et al.~\cite{ChanCCMS2019} show that DO can improve asymptotic complexity over full obliviousness with constant $\eps$ and negligible $\delta$. For example, they give stable one-bit sorting in $O(T\log\log T)$ time with constant private storage in the balls-and-bins model, improving on the $\Omega(T\log T)$ bound for full obliviousness. Whether these gains extend to integer priority queues with polylogarithmic private space remains open. Therefore, we consider the following question with constant $\eps$ and negligible $\delta$.
\begin{center}
\em Can any $(\eps,\delta)$-DO integer priority queue achieve $o(\log T)$ cost with polylogarithmic private space?
\end{center}

\subsection{Our Contributions}
\label{sec:contributions}

We answer this question affirmatively by constructing an exact DO heap with sublogarithmic total work and bandwidth in the balls-and-bins model. We also prove a matching lower bound. Our upper bound and its optimality in this model can be summarized in the following theorem, where $T\ge2$ is a public upper bound on the length of the input operation sequence.

\begin{theorem}[Upper bound, informal]
\label{thm:informal-main}
Let $\eps=\Theta(1)$ and $\delta=\exp(-\Theta(\log^2 T))$. For any constant $c\ge1$, there exists an online priority queue $\mathcal{Q}$ that supports $\mathcal{Q}.\Insert$ and $\mathcal{Q}.\Extract$ operations with integer priorities bounded by $O(T^c)$, satisfies $(\eps,\delta)$-differential obliviousness, incurs $O(\log T/\log\log T)$ amortized total work in expectation, and requires only $\polylog(T)$ private words.
\end{theorem}

Further, we provide a matching lower bound, stated in the following theorem.

\begin{theorem}[Lower bound, informal]
\label{thm:informal-lower}
For $\eps=O(1)$ and $\delta=o(1)$, any exact online $(\eps,\delta)$-DO priority queue in the balls-and-bins model, with integer priorities in $[T]$, $w=\Theta(\log T)$-bit words, and $\polylog(T)$ private words, requires $\Omega(\log T/\log\log T)$ expected amortized probes per operation.
\end{theorem}

Table~\ref{tab:pq-comparison} compares the operation costs in the comparison and integer settings, where non-oblivious costs are measured on the ordinary RAM, and oblivious constructions may use polylogarithmic private storage.

\begin{table}[!hb]
\makeatletter
\long\def\@makecaption#1#2{%
  \vskip\abovecaptionskip
  \sbox\@tempboxa{#1. #2}%
  \ifdim\wd\@tempboxa>\hsize
    #1. #2\par
  \else
    \global\@minipagefalse
    \hb@xt@\hsize{\hfil\box\@tempboxa\hfil}%
  \fi
  \vskip\belowcaptionskip}
\makeatother
\centering
\begin{tabularx}{\textwidth}{l>{\centering\arraybackslash}X>{\centering\arraybackslash}X@{}}
\toprule
{\bf Privacy guarantee} & {\bf Comparison-based PQ} & {\bf Integer PQ} \\
\midrule
Non-oblivious & $\Theta(\log T)$ & $\Theta(1)$~\cite{Thorup2007} \\
Fully oblivious & $\Theta(\log T)$~\cite{Shi2020PathObliviousHeap,JafargholiLarsenSimkin2021,JacobLarsenNielsen2019} & $\Theta(\log T)$~\cite{Shi2020PathObliviousHeap,JafargholiLarsenSimkin2021,JacobLarsenNielsen2019} \\
Differentially oblivious & $\Theta(\log T)$ & $\Theta(\frac{\log T}{\log\log T})$ (our work) \\
\bottomrule
\end{tabularx}
\caption{Expected amortized cost of priority queues in the balls-and-bins model. The fully oblivious upper bounds allow inverse-polynomial correctness error~\cite{Shi2020PathObliviousHeap,JafargholiLarsenSimkin2021}, while our construction is exact.}
\label{tab:pq-comparison}
\end{table}

\section{Preliminaries}
\label{sec:prelim}



A \emph{priority queue} stores records of the form $(p,v)$, where $p$ is a priority and $v$ is its payload. Given a priority queue $\mathcal{Q}$, it supports the following operations:
\begin{itemize}[leftmargin=*]
\item $\mathcal{Q}.\Insert(p,v)$ adds the record $(p,v)$ to $\mathcal{Q}$.
\item $\mathcal{Q}.\Extract()$ removes and returns a minimum-priority record, or returns $\bot$ if $\mathcal{Q}$ is empty. Ties are resolved by insertion order.
\end{itemize}
Operations arrive online, and $T\ge 2$ is a public upper bound on their number. We write $[n]=\{1,\ldots,n\}$ for any $n\in\N_+$ and assume priorities lie in $[U]$ for a public $U=T^{O(1)}$. To break ties, we assign each inserted record a key $(p,t)$, where $p$ is its priority and $t$ is the index of its insertion operation. Keys are distinct and are compared lexicographically.

By concatenating fixed-length binary encodings of $p$ and $t$, we identify keys with integers in the public universe $[V]$, where $V=2^{\lceil\log(U+1)\rceil+\lceil\log(T+1)\rceil}=T^{O(1)}$. This encoding preserves lexicographic order. All constructions below use these encoded keys $k\in [V]$.

\subsection{Computation model and access patterns}
\label{sec:balls-and-bins}

Let $f\colon\mathcal{X}\to\mathcal{Y}$ be a (randomized) RAM-implemented algorithm. For an input $x\in\mathcal{X}$, we use $\view^f(x)$ to denote the \emph{access pattern}, i.e., the chronological sequence of observable physical memory accesses made during its execution. Each access records its address and read/write type.

Given $f$, the specific content of $\view^f(\cdot)$ depends on the observation surface. We use the client--server setting widely adopted in oblivious RAMs~\cite{GoldreichOstrovsky1996,StefanovEtAl2013PathORAM}, in which a client with private memory accesses a passive server. Under this setting, only the server-side accesses are observable to the adversary, while the client-side accesses remain hidden. For an operation sequence $x$ of length $n$, we write $\view^f(x)=(V_1,\ldots,V_n)$, where $V_i$ lists the server accesses made during the $i$-th operation and may be empty. This grouping reveals operation boundaries. Operation types, private arguments, outputs, and elapsed time are not included in the trace.

Throughout the paper, we adopt the \emph{balls-and-bins model}~\cite{ChanCCMS2019,ChanChungMaggsShi2022}, and assume that the payloads of all priority queue nodes are opaque balls that may be stored, moved, copied, or discarded, but cannot be split or jointly encoded. For storage, we assume the word size to be $w=\Theta(\log T)$-bit, and each record, including its key, opaque payload, and metadata, occupies $O(1)$ words. Computations on keys and metadata follow the word-RAM model with constant-time multiplication~\cite{FredmanWillard1993,PatrascuThorup2014}.

\paragraph{Cost model}
Based on the settings above, reading or writing any single server word costs one \emph{probe}. The \emph{bandwidth cost} is the number of words transferred between the client and the server and is within a constant factor of the probe count. The \emph{total work} counts all client word operations and server probes. Private space counts all persistent and temporary client storage, including input/output registers and retained answers. Encryption and decryption of server-side data are implicit and omitted from our description. The primitive instructions are specified in Appendix~\ref{app:word-ram}.

\subsection{Obliviousness and differential obliviousness}
\label{sec:privacy-definition}
Given the observation model defined previously, \emph{(full) obliviousness} requires the distributions of $\view^f(x)$ and $\view^f(x')$ to be statistically indistinguishable for any two inputs $x,x'\in\mathcal{X}$ with the same public parameters. The formal definition is as follows.
\begin{definition}[Obliviousness~\cite{GoldreichOstrovsky1996}]
  \label{def: o}
  Given a security parameter $\lambda$, a randomized RAM-implemented algorithm $f\colon\mathcal{X}\to\mathcal{Y}$ is {\em (statistically) oblivious} if for every pair of inputs $x,x'\in\mathcal{X}$ with the same public parameters,
  \begin{align*}
    \Delta(\view^f(x),\view^f(x')) \leq \negl(\lambda),
  \end{align*}
  where $\Delta(\xi,\zeta)=\frac{1}{2}\sum_z |\Prb[\xi=z]-\Prb[\zeta=z]|$ denotes the statistical distance, and $\negl$ is an unspecified negligible function, i.e., $\negl(\lambda)=o(\lambda^{-c})$ holds for every constant $c$.
\end{definition} 

\emph{Differential obliviousness}~\cite{ChanCCMS2019,ChanChungMaggsShi2022} relaxes this requirement by replacing the statistical indistinguishability with {\em differential privacy}~\cite{Dwork06D} on the access patterns of neighboring inputs. The neighboring relation specifies which changes to an input are protected.

\begin{definition}[Differential obliviousness~\cite{Dwork06D,ChanCCMS2019}]
\label{def:do}
Let $\neigh$ be a symmetric neighboring relation on $\mathcal{X}$. For $\eps\ge0$ and $\delta\in[0,1]$, a RAM-implemented algorithm $f\colon\mathcal{X}\to\mathcal{Y}$ is \emph{$(\eps,\delta)$-differentially oblivious ($(\eps,\delta)$-DO for short)} if $\view^f(\cdot)$ satisfies $(\eps,\delta)$-differential privacy ($(\eps,\delta)$-DP for short), i.e., for every pair $x\neigh x'$ and every measurable set $\mathcal{S}$ of finite access patterns,
\begin{align*}
\Prb[\view^f(x)\in\mathcal{S}]
\le e^\eps\cdot \Prb[\view^f(x')\in\mathcal{S}]+\delta.
\end{align*}
\end{definition}

In Definition~\ref{def:do}, probabilities are over the internal randomness of $f$. Similar to DP, for $(\eps,\delta)$-DO, smaller $\eps$ and $\delta$ provide stronger privacy guarantees.

Within our scope, for a priority queue, an input is an operation sequence $\boldsymbol{\op}=(\op_1,\op_2,\ldots,\op_n)$, where $n\le T$ and each $\op_i$ is of the form $\Insert(p,v)$ or $\Extract()$. We write $|\boldsymbol{\op}|$ for the length of $\boldsymbol{\op}$. Without loss of generality, we assume that any priority queue $\mathcal{Q}$ starts with an empty initial state. Nonempty initial contents can be represented by an insertion prefix in an order consistent with tie breaking. Then two operation sequences $\boldsymbol{\op}$ and $\boldsymbol{\op}'$ with the same public parameters and equal length $n$ are neighboring, written $\boldsymbol{\op}\neigh\boldsymbol{\op}'$, if there is at most one index $i\in[n]$ with $\op_i\ne\op_i'$. The differing operation may change its type and private fields.

Finally, we introduce two properties of DP, which are used in our analysis.

\begin{theorem}[Sequential composition~\cite{DworkRoth2014}]
Let $M_1,M_2\colon\mathcal{X}\to\mathcal{Y}$ satisfy $(\eps_1,\delta_1)$-DP and $(\eps_2,\delta_2)$-DP, respectively, under the same neighboring relation using independent randomness. Then their joint mechanism $x\mapsto(M_1(x),M_2(x))$ satisfies $(\eps_1+\eps_2,\delta_1+\delta_2)$-DP.
\end{theorem}

\begin{theorem}[Post-processing~\cite{DworkRoth2014}]
Let $f\colon\mathcal{X}\to\mathcal{Y}$ satisfy $(\eps,\delta)$-DP. For any fixed function $g\colon\mathcal{Y}\to\mathcal{Z}$, the composition $g\circ f$ also satisfies $(\eps,\delta)$-DP.
\end{theorem}

\section{Overview}
\label{sec:overview}

For the upper bound, i.e., the construction of DO priority queue, we consider a preprocessing-based scheme. Specifically, our main idea is to prepare a small private working set for each epoch of $B=\polylog(T)$ operations. We first consider public operation types. At each epoch's start, the working set contains the $B$ live records with the smallest keys, or all live records if fewer remain. These records, together with new insertions, suffice because the epoch contains at most $B$ extractions. Records from completed epochs remain on the server even after extraction, with their deletion state tracked separately.

The server organizes these historical records into a hierarchy of segments, each sorted by key and partitioned into bins of $O(B)$ slots. A segment's capacity is the number of operations in the epoch interval it covers. Capacities grow by a factor $\ell=\Theta(\log T)$ between levels, which ensures the number of levels to be $O(\log T/\log\log T)$. At an epoch boundary, that epoch's insertions form a new segment, and a public base-$\ell$ carry rule determines which segments merge into a larger segment. Each level has at most $\ell-1$ segments between epoch boundaries, and each operation position belongs to at most one created segment per level.

On the client, new insertions enter a stash of at most $B$ records, each with a live bit. Each segment has a private local set containing its prefetched records that are still live. An integer index $\mathcal{I}$ maintains the keys of all live stash records and all records in the local sets. Under our word-RAM assumptions, this polylogarithmic-size index supports insertion, deletion, and minimum selection in constant time~\cite{PatrascuThorup2014}. An extraction removes the minimum record from the index and clears its stash live bit or increments its segment's private extraction counter. Subsequent extractions from a segment remove a prefix of the records that were live at its creation, so this counter suffices to track those deletions while preserving the original records.

To prepare these local sets for the next epoch, we seek a key cutoff whose prefix contains enough live records. The cutoff and bin boundaries depend on private counts, so accesses based directly on those counts would leak information. We instead construct a differentially private summary for each segment, estimating the number of its original records with key at most $x$ (Lemmas~\ref{lem:bounded-noise} and~\ref{lem:static-summary}). These include records extracted before the segment was created. Changing one insertion can change later extraction results, but only segments containing that insertion have different multisets of inserted keys for summary construction. The old summaries determine new bin boundaries during a merge. The new summary is computed afresh from the original records, maintaining the same error guarantee across levels.

To obtain estimates of live counts from these historical counts, we maintain a separate segment tree over the key universe on the server. It stores an estimate $\widehat{Q}(x)$ of the number of live records with key at most $x$. An extraction removes one record from every nonempty key prefix, so its effect on all prefix counts is the pointwise map $v\mapsto\max\{v-1,0\}$, independent of the extracted key. The tree applies this update lazily at its root and incorporates changes to the segment summaries at epoch boundaries. Summing the current segment summaries estimates the total historical insertion count at each key threshold. The error bound also accounts for the at most $B$ stash insertions not yet represented. The reflection lemma transfers this historical-count error bound to the live-count estimate with only a constant-factor loss, independent of the number of extractions (Lemma~\ref{lem:reflection}).

We choose $B$ large enough to absorb the aggregate summary error, so the live-count error is $O(B)$. Searching for the first key at which $\widehat{Q}$ crosses $B$ plus an error margin supplies a prefix containing at least $B$ live records and bounds the excess by $O(B)$. If no crossing exists, the total live count is $O(B)$, and the cutoff is the end of the key range. Within each segment, the client extends the previously read bin prefix to cover the cutoff and retains live copies in the local set. Reading the complete boundary bin adds at most $O(B)$ records. Between extensions, a segment's local set only shrinks as records are extracted. Thus each local set has size $O(B)$, and the prefetched records suffice for exact answers throughout the next epoch (Lemma~\ref{lem:local-set}).

It remains to construct these summaries with the required error and work guarantees. The summary mechanism adds bounded noise to dyadic counts, retains nodes whose noisy counts exceed a threshold, and estimates the others by zero. The threshold and noise bound control each node's error, so dyadic prefix decomposition gives a uniform error bound on every execution. Bounded noise lets a compressed-tree scan skip nodes that cannot be retained. With our parameter choice, processing the remaining candidates takes expected work linear in the segment capacity. Quantization produces a short weighted list of positions in the key universe, keeping the resulting segment-tree updates within the same linear work bound.

To bound the joint privacy loss, we use the fact that each insertion contributes to at most one fresh summary per level. Allocating the privacy budget across levels bounds the joint privacy loss by composition. During summary generation, the sequence of whole-bin reads and padded writes is determined by public capacities and the previously released layout. The remaining accesses, including merging and prefetching, depend only on the released noisy nodes and public operation types. Thus the complete access pattern inherits the privacy guarantee of the joint releases.

The hierarchy also controls the total work. Constructing a segment takes expected work linear in its capacity (Lemmas~\ref{lem:merge} and~\ref{lem:streaming-summary}). Merges process bins in private memory, where integer indexing permits linear-time sorting of each bin. Each operation position is charged for at most one construction per level, and each bin is prefetched at most once during its segment's lifetime. Our choice of $B$ also absorbs searches at epoch and bin boundaries. These charges give constant expected amortized work per level.

Once the working set is prepared, calls use the private index, with constant-cost root updates for extractions. Together with epoch maintenance, this gives $O(\log T/\log\log T)$ expected amortized total work per operation. To remove the assumption of public operation types, a wrapper replaces each external operation with a fixed sequence of two insertions and one extraction. Auxiliary records preserve the external queue's answers, with only constant-factor overhead.

For the lower bound, we compare the entropy of the output permutation with the number of permutations compatible with a fixed access trace. Consider $n$ insertions of distinct opaque payloads with independent uniform priorities in $[n]$, followed by $n$ extractions, where $2n\le T$. An exact queue returns a permutation with entropy $n\log n-O(n)$. For $\eps=O(1)$, differential privacy bounds the information that the trace $\view$ reveals about the priorities by $O(n+n\delta\log n)$. Thus, for $\delta=o(1)$, the conditional entropy of the output permutation given $\view$ remains $\Omega(n\log n)$ (Lemma~\ref{lem:lb-entropy}).

With $M$ private words, a fixed trace of $q$ probes supports at most $(M+2)^{O(q+n)}$ output permutations. The opaque-payload restriction permits a routing description using $O(\log(M+2))$ bits per probe, input arrival, or output (Lemma~\ref{lem:lb-routing}). Comparing this counting bound with the residual entropy yields $\Omega(n\log n/\log\log n)$ expected probes for $M=\polylog(n)$. Averaging over the input distribution gives a fixed history with at least this expected cost. Taking $n=\lfloor T/2\rfloor$, when the priority domain contains $[n]$, gives the matching $\Omega(\log T/\log\log T)$ expected amortized probe lower bound.

\section{DO Priority Queue Construction}
\label{sec:construction}

This section introduces the construction of a DO priority queue. As stated in Section~\ref{sec:prelim}, we allow a client with $\polylog(T)$ storage, with accesses to that storage hidden from the adversary. We first construct a priority queue with public operation types, which we call the \emph{core}, supporting $\Insert$ and $\Extract$ operations. A \emph{wrapper} then hides operation types by translating each external operation into the fixed sequence $\Insert,\Insert,\Extract$ of core operations.

We first state the main guarantee.

\begin{theorem}[DO priority queue]
\label{thm:main}
Fix constants $c,\eps>0$ and $T\in\N_+$. Let $\ell=\lceil\log T\rceil$ and $\delta_T=2^{-\ell^2}$. There is an exact online $(\eps,\delta_T)$-DO priority queue $\mathcal{Q}$, where the priority range is $[T^c]$. For every fixed sequence of $t\le T$ calls to $\mathcal{Q}.\Insert$ or $\mathcal{Q}.\Extract$, it requires $\log^{O(1)}T$ private words, and incurs expected total work of $O(t\cdot \log T/\log\log T)$.
\end{theorem}

Sections~\ref{sec:layout}--\ref{sec:segments} describe the core's data structures, operations, and summary construction. Section~\ref{sec:wrapper} defines the wrapper that hides operation types. Section~\ref{sec:analysis} proves Theorem~\ref{thm:main} and gives explicit dependence on the privacy parameters.

\subsection{Data Structure and Parameters}
\label{sec:layout}


As described in Section~\ref{sec:overview}, we divide the core operation sequence into \emph{epochs} of $B=\polylog(T)$ consecutive operations. At the start of each epoch, the client prefetches a set containing the $B$ live records with the smallest keys, or all live records if fewer remain. Since an epoch contains at most $B$ extractions, these records together with new insertions suffice to answer all operations in the epoch. We therefore keep current insertions in a stash, store earlier insertions in a hierarchy of sorted segments, and retain prefetched live records in local sets. The table below summarizes the five components of this construction, their locations, and their roles.

\begin{center}
\renewcommand{\arraystretch}{1.15}
\begin{tabularx}{\linewidth}{lll}
\toprule
{\bf Component} & {\bf Location} & {\bf Role} \\
\midrule
Stash & Client & Store all records inserted during the current epoch \\
Segments & Server & Store all original records from completed epochs \\
Local sets & Client & Retain copies of prefetched records as candidates for the current epoch \\
Integer index & Client & Indexes the keys of stash and local sets for minimum selections \\
Segment tree & Server & Maintains an estimate $\widehat{Q}(x)$ of the number of live records with key $\leq x$ \\
\bottomrule
\end{tabularx}
\end{center}

\paragraph{Stash.} The \emph{stash} is an array of at most $B$ entries, one for each insertion in the current epoch, of the form $\mathsf{stash}[i]=(k_i,v_i,b_i)$. Here $k_i\in[V]$ is the encoded insertion key, $v_i$ is its payload, and $b_i\in\{0,1\}$ denotes its {\em live bit}, i.e., it equals $1$ if and only if the record has not been extracted. Extracted entries remain in the stash until the epoch ends.

\paragraph{Segment.} The {\em sorted segments}, or simply {\em segments}, are organized in a hierarchical structure. Let $d=\lceil \log T/\log\ell \rceil$, where $\ell=\lceil \log T\rceil$ is the growth factor from Theorem~\ref{thm:main}. The hierarchy has $d+1=O(1+\log T/\log\log T)$ levels, labeled by $0,\ldots,d$. At the start of each epoch, each level contains at most $\ell-1$ segments, and the epoch intervals of all segments are pairwise disjoint. Thus $K=1+(\ell-1)(d+1)$ is a public upper bound on the number of segments, including a temporary level-$0$ segment created at an epoch boundary. A level-$j$ segment stores all records inserted during an interval of $\ell^j$ consecutive completed epochs, sorted by key. We call these its \emph{original records}, including those that have already been extracted. Its capacity bound is $m_0=B\cdot\ell^j$, the number of core calls in the interval, and its record count $m\le m_0$ is the total number of insertions in that interval. Both $m$ and $m_0$ are public, since epoch boundaries and operation types are public.

\paragraph{Bin.} Every level-$j$ segment $r$ is further partitioned into $\ell^j$ {\em bins}, each with $2B$ physical slots. Its {\em bin directory} is an array of key boundaries
\begin{align*}
0=q_{r,0}\le q_{r,1}\le\cdots\le q_{r,\ell^j}=V.
\end{align*}
For each $i\in[\ell^j]$, bin $i$ stores exactly the original records of $r$ whose keys lie in $(q_{r,i-1},q_{r,i}]$, in increasing key order. The remaining slots are empty. In particular, when $q_{r,i-1}=q_{r,i}$, all $2B$ slots are empty. In addition to the record $(k,v)$, each nonempty slot stores a deletion bit $\mathsf{dead}$, which equals $1$ if and only if the record had been extracted before $r$ was created. Each slot also carries a bit $\mathsf{pad}$, equal to $1$ for empty slots and $0$ otherwise. The remaining contents of an empty slot are ignored.

\begin{figure}[!t]
\centering
\begin{tikzpicture}[
  x=1cm,y=1cm,
  font=\fontsize{9}{11}\selectfont,
  heapbox/.style={draw=black!65,line width=.55pt,rounded corners=1.5pt,
    fill=white,align=center,inner sep=4pt},
  heaparrow/.style={-{Stealth[length=2mm,width=1.5mm]},line width=.7pt},
  flowlabel/.style={font=\fontsize{8}{9.5}\selectfont,align=center,fill=white,inner sep=1pt}
]
\newcommand{\heapsegment}[5]{%
  \begin{scope}[shift={(#1,#2)}]
    \pgfmathsetmacro{\segmentwidth}{#3*.64}
    \draw[draw=black!75,line width=.65pt,rounded corners=1pt]
      (-.07,-.30) rectangle ({\segmentwidth+.07},.30);
    \foreach \slot in {1,...,#3}{
      \ifnum\slot>#4\relax
        \def\binfill{white}
      \else
        \def\binfill{blue!22}
      \fi
      \draw[draw=black!55,fill=\binfill,line width=.4pt]
        ({(\slot-1)*.64+.035},-.22)
        rectangle ({\slot*.64-.035},.22);
    }
    \node[font=\fontsize{8}{9}\selectfont] at ({\segmentwidth/2},.47) {$#5$};
  \end{scope}%
}

\draw[draw=black!50,fill=black!2,rounded corners=3pt]
  (0,1.5) rectangle (6.15,6.15);
\draw[draw=black!50,fill=black!2,rounded corners=3pt]
  (8.05,1.5) rectangle (16.45,6.15);
\node[font=\small\bfseries] at (3.075,5.85) {Client};
\node[font=\small\bfseries] at (12.25,5.85) {Server};

\node[heapbox,minimum size=1cm,inner sep=2pt] (stash) at (3.8,4.65)
  {$\mathsf{stash}$};
\node[font=\fontsize{9}{11}\selectfont\bfseries] at (3.8,5.4) {Stash};
\node[heapbox,minimum size=1cm,inner sep=2pt] (index) at (1.35,3.65)
  {$\mathcal{I}$};
\node[font=\fontsize{9}{11}\selectfont\bfseries] at (1.35,4.4) {Integer index};
\node[font=\fontsize{9}{11}\selectfont\bfseries] at (3.8,3.15) {Local sets};
\foreach \i/\x/\nbins in {1/2.2/1,2/3.0/1,3/3.8/1,4/4.6/2,5/5.4/3}{
  \node[heapbox,minimum size=.7cm,inner sep=2pt] (localset\i) at (\x,2.5) {};
  \node[inner sep=0pt] at (\x,2.63) {$\mathcal{C}_{r_\i}$};
  \foreach \slot in {1,...,\nbins}{
    \draw[draw=black!55,fill=blue!22,line width=.4pt]
      ({\x+(\slot-(\nbins+1)/2)*.18-.07},2.25)
      rectangle ({\x+(\slot-(\nbins+1)/2)*.18+.07},2.39);
  }
}
\draw[draw=black!65,line width=.7pt] (index.east) -- (stash.west);
\draw[draw=black!65,line width=.7pt]
  (index.south) -- (1.35,1.75) -- (5.4,1.75);
\foreach \i/\x in {1/2.2,2/3.0,3/3.8,4/4.6,5/5.4}{
  \draw[draw=black!65,line width=.7pt] (localset\i.south) -- (\x,1.75);
}

\node[font=\fontsize{9}{10.5}\selectfont] at (12.25,5.3)
  {{\bf\boldmath Segment hierarchy ($\ell=3$)}};
\node[anchor=west,font=\fontsize{8}{9.5}\selectfont] at (8.3,4.55) {Level $0$};
\node[anchor=west,font=\fontsize{8}{9.5}\selectfont] at (8.3,3.6) {Level $1$};
\node[anchor=west,font=\fontsize{8}{9.5}\selectfont] at (8.3,2.65) {Level $2$};
\heapsegment{9.55}{4.55}{1}{1}{r_1}
\heapsegment{11.15}{4.55}{1}{1}{r_2}
\heapsegment{9.55}{3.6}{3}{1}{r_3}
\heapsegment{12.3}{3.6}{3}{2}{r_4}
\heapsegment{9.55}{2.65}{9}{3}{r_5}

\draw[draw=black!55,fill=blue!22] (9.28,1.8) rectangle (9.55,2.0);
\node[anchor=west,font=\fontsize{8}{9.5}\selectfont] at (9.7,1.9)
  {Prefetched bins};

\draw[heaparrow,draw=orange!75!black] (stash.east) -- (8.05,4.65)
  node[pos=.75,above=3pt,flowlabel] {Epoch end};
\draw[heaparrow,draw=blue!65!black] (8.05,2.5) -- (5.9,2.5)
  node[midway,above=3pt,flowlabel] {Prefetch};
\end{tikzpicture}
\caption{Client--server layout with $\ell=3$. The stash and local sets are stored on the client, and $\mathcal{I}$ provides an index for them. Segments are stored on the server, where each small rectangle is a bin with $2B$ slots. Shaded bins form the prefetched prefix of each segment.}
\label{fig:doheap-layout}
\end{figure}
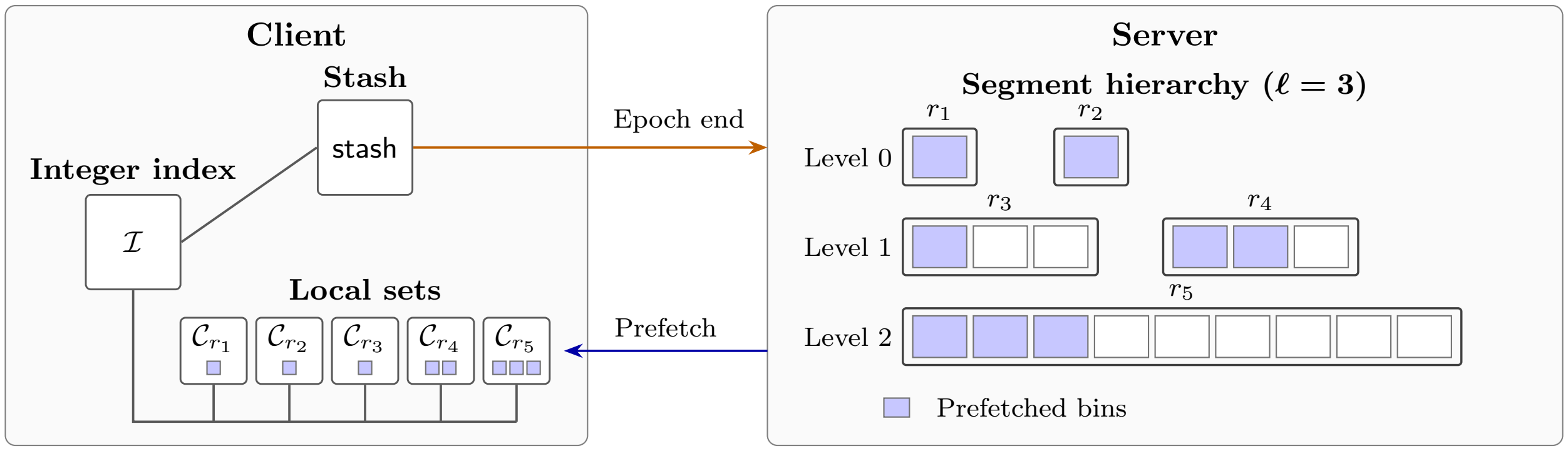

\paragraph{Local sets and integer index.}
For each segment $r$, the client maintains a \emph{local set} $\mathcal{C}_r$ of copies of the live records in a prefix of its bins. As visualized in Figure~\ref{fig:doheap-layout}, $\mathcal{C}_r$ contains copies of the live records in the shaded bins of segment $r$. Reading and extending these prefixes to prepare for an epoch is called \emph{prefetching}. Let $f_r$ denote the number of initial bins of $r$ that have been prefetched. Let $\rho_{r,1},\rho_{r,2},\ldots$ denote the records that were live when segment $r$ was created, listed in increasing key order. The counter $c_r$ counts records extracted from $r$ since its creation, and $\seen_r$ counts the records $\rho_{r,k}$ encountered in the read prefix. After each operation, the local set satisfies
\begin{align*}
\mathcal{C}_r=\{\rho_{r,k}\mid c_r<k\le\seen_r\}.
\end{align*}
The frontiers $f_r$ are public, while the counters $c_r$ and $\seen_r$ remain private. At the start of each epoch, prefetching ensures that the union of the local sets contains the smallest $B$ live records inserted before that epoch, or all such records if fewer remain.

An integer set $\mathcal{I}$ indexes the keys of all live stash records and all records in $\bigcup_r\mathcal{C}_r$. Each key has a pointer to its record and an owner, either $\mathsf{stash}$ or the corresponding segment $r$. There are $O(KB)=w^{O(1)}$ entries under the parameter choices of Theorem~\ref{thm:main}, so the following lemma gives constant-time insertion, deletion, and minimum selection.

\begin{lemma}[Integer sets~\cite{PatrascuThorup2014}]
\label{lem:small-integer-sets}
On a word-RAM with $w$-bit words and constant-time multiplication, there exists a data structure that maintains a dynamic set of $n$ distinct word keys using $O(n)$ words and supports insertion, deletion, and selection in $O(1+\log n/\log w)$ time per operation. In particular, when $n=w^{O(1)}$, these operations, including minimum selection, take constant time.
\end{lemma}

The integer set's supporting constants require $O(\log w)$ setup work~\cite{PatrascuThorup2014}. Section~\ref{sec:probe-accounting} accounts for this setup and all temporary storage.

\paragraph{Segment tree.}
To determine which bin prefixes to prefetch, we construct an implicit segment tree over $[V]$ on the server, allocating nodes as needed. For each $x\in[V]$, the tree maintains an estimate $\widehat{Q}(x)$ of the number of live records with key at most $x$. To maintain this estimate, we associate each segment with a differentially private summary of its original records.

For the main result, fix core privacy parameters $\eta>0$ and $0<\delta<1$, where $\eta$ is constant and $\log(1/\delta)=O(\ell^2)$.
Let $\mathcal{R}_r$ denote the set of records inserted during the epochs covered by segment $r$, regardless of whether they have subsequently been extracted. Padding slots are excluded, and $|\mathcal{R}_r|=m$. For all $x\in\{0,\ldots,V\}$, the number of these records with key at most $x$ is $F_r(x)=\sum_{(k,v)\in\mathcal{R}_r}\one_{k\le x}$. The procedure $\mathsf{BuildSummary}(r)$ ignores deletion bits and uses fresh randomness to produce a differentially private vector $\mathcal{S}_r$ of noisy counts for selected key intervals. From this vector and public parameters, it derives a weighted list with positions $z_{r,1}\le z_{r,2}\le\cdots$ in $[V]$ and nonnegative weights $\omega_{r,i}$. This list represents the cumulative summary function
\begin{align}
S_r(x)=\sum_{z_{r,i}\le x}\omega_{r,i}.
\label{eq:stored-summary}
\end{align}
Here $S_r(x)$ is a differentially private estimate of the true prefix count $F_r(x)$.
The count quantization step is $s=\ell^6$. The list has $\lfloor m/s\rfloor$ entries of weight $s$ and a final entry at $V$ of weight $m-s\lfloor m/s\rfloor$, which may be zero. Positions may repeat.

The summary construction uses a conceptual complete binary tree obtained by recursively bisecting $[V]$ down to individual keys. This \emph{dyadic count tree} has $h=1+\log V$ levels and is distinct from the segment tree maintaining $\widehat{Q}$. Choose fixed positive integers $\kappa\ge1+e^\eta$ and $c_\eta\ge1/\eta$. Let $L$ be the least power of two at least $\max\{2,4c_\eta h(d+1)\}$, and set
\begin{align*}
R_0=\max\left\{\ell^2,\left\lceil\log(1/\delta)\right\rceil+\left\lceil\log \bigl(4\kappa h(d+1)\bigr)\right\rceil\right\},\quad
H=LR_0.
\end{align*}
Here $L$ determines the geometric sampling probability $1/L$ in Section~\ref{sec:segments}, and $H$ caps each geometric draw. The ratio $R_0=H/L$ controls the probability that an uncapped draw exceeds $H$. Define the public uniform error bound for one segment summary by
\begin{align}
Z=s+h(3H+1).
\label{eq:summary-error}
\end{align}
On every execution, $S_r$ is monotone, $S_r(V)=m$, and $\|S_r-F_r\|_\infty\le Z$. We choose the epoch length, which also determines the bin size, as
\begin{align}
B=\left\lceil\ell^{12}+4KZ+4\right\rceil.
\label{eq:buffer-parameters}
\end{align}
All these parameters depend only on public inputs. Sections~\ref{sec:segments} and~\ref{sec:probe-accounting} establish the summary guarantees and resource bounds for these choices.

Each vector $\mathcal{S}_r$ is $\bigl(\frac{\eta}{d+1},\frac{\delta}{d+1}\bigr)$-DP under replacement of one original record. Its distribution depends only on the original records and fresh randomness. Both $S_r$ and the access pattern of its construction are postprocessing of $\mathcal{S}_r$, the existing layout determined by earlier summary vectors, and public parameters. Section~\ref{sec:segments} implements this interface using expected $O(m_0)$ work, $O(m_0)$ probes on every execution, and $O(B)$ additional private words.

We call the segments currently installed in the hierarchy \emph{committed segments}. The segment tree is updated using their summaries and the public sequence of operation types. Interval maxima and lazy updates support the choice of a key threshold that determines how many initial bins of each segment to read. Section~\ref{sec:operations} defines these updates and searches.
The public error bound for the cumulative insertion counts used in these updates is
\begin{align}
E=KZ+B.
\label{eq:prefetch-parameters}
\end{align}
Here $KZ$ accounts for the segment summaries, and $B$ accounts for the stash records, which are not yet included in those summaries.

Each level has a separate server address range, and new segments use fresh addresses. All structures are initialized only when used.

\subsection{Queue Operations}
\label{sec:operations}

At the start of each epoch, $\bigcup_r\mathcal{C}_r$ contains the $B$ live records with the smallest keys, or all live records if fewer remain. The stash retains every insertion during that epoch. These records suffice for all $B$ calls, since at most $B$ old records can be extracted. Lemma~\ref{lem:local-set} proves this property and gives $|\mathcal{C}_r|=O(B)$ for every $r$.

To select the bin prefixes needed for this coverage, we maintain $\widehat{Q}$ using the summaries $S_r$ and extraction counts. We describe these updates first, followed by insertion and extraction using $\mathcal{I}$, prefetching into $\mathcal{C}_r$, and updates to the segment hierarchy at epoch boundaries.

\paragraph{Prefix counts.}
Let $A(x)$ count all records inserted so far with key at most $x$, including records already extracted and those in the current stash. We estimate this count by $\widehat{A}(x)=\sum_r S_r(x)$, where $r$ ranges over the currently committed segments.
An extraction decreases a positive live prefix by one and leaves an empty prefix at zero. We apply the same reflection to the estimated insertion increments. An event index $t$ includes both calls and replacements of committed segments. Each replacement is a single event. Let $N_t$ count all extraction calls through event $t$, including empty ones, and write $A_t(x)$, $\widehat{A}_t(x)$, and $\widehat{Q}_t(x)$ for the corresponding values after that event. With $[z]_+=\max\{z,0\}$, $N_0=0$, and $\widehat{Q}_0=0$, we maintain
\begin{align*}
\widehat{Q}_t(x)=\bigl[\widehat{Q}_{t-1}(x)+\widehat{A}_t(x)-\widehat{A}_{t-1}(x)-(N_t-N_{t-1})\bigr]_+.
\end{align*}
Lemmas~\ref{lem:reflection} and~\ref{lem:cutoff} show that the first crossing of $B+2E$ gives a prefix containing enough live records with a bounded excess.

The pointwise updates and threshold searches below are implemented by the segment tree. A suffix addition adds $\omega$ at all coordinates at least $z$. A root reflection applies $v\mapsto[v+\lambda]_+$ lazily to all coordinates. A threshold search finds the smallest coordinate with value at least $\tau$, or returns $V$ if none exists. Suffix addition and search cost $O(\log V)$ probes and word operations. Root reflection costs $O(1)$. Search uses interval maxima and does not require $\widehat{Q}$ to be monotone in $x$.

An extraction uses root reflection with $\lambda=-1$. A stash insertion leaves $\widehat{A}$ unchanged. At commit, we remove the old summaries and add the new summary using suffix additions, then truncate negative values to zero by root reflection with $\lambda=0$. This order implements a single replacement event. A temporary level-$0$ segment is not separately committed when it is merged with other segments. The lazy transformations have the form $v\mapsto\max\{v+u,z\}$ and are closed under composition. Pure additions use a flag for $z=-\infty$. Appendix~\ref{app:word-ram} gives the propagation details. Only a root pointer, an allocation counter, and an $O(\log V)$ traversal stack remain private. Accesses and allocations depend only on released summaries and public types.

\paragraph{Insertions and extractions.}
An insertion appends its original fields and a live bit to the stash and inserts its key into $\mathcal{I}$. An extraction removes the minimum indexed record, or returns $\bot$ if the index is empty. If the record belongs to the stash, its live bit is cleared. Otherwise, it is removed from its local set $\mathcal{C}_r$ and $c_r$ increases by one. The original fields remain available for future summaries. After $B$ calls, $\mathsf{CloseEpoch}$ (Algorithm~\ref{alg:close}) commits the next segment and $\mathsf{Refresh}$ restores the coverage property.

An index entry points to a record in the stash or in some $\mathcal{C}_r$. Adding a record to $\mathcal{C}_r$ creates a private copy with owner $r$. Representing each $\mathcal{C}_r$ as a linked list supports both operations in $O(1)$ local work, without changing server records. The original priority is the high-order component of the encoded key and is recovered by a bit shift. The counters $i$ and $t_{\mathrm{call}}$ count calls in the current epoch and in the complete history, respectively. Both counters start at zero.

\begin{algorithm}[!htb]
\SetAlgoCaptionSeparator{.}
\caption{CoreCall$(\mathsf{op})$}
\label{alg:online}
\KwIn{One core operation, with private fields $(p,v)$ for an insertion}
\KwOut{The exact private answer for an extraction}
$t_{\mathrm{call}}\gets t_{\mathrm{call}}+1$\;
\eIf{$\mathsf{op}=\Insert(p,v)$}{
  Append a live record with encoded key $(p,t_{\mathrm{call}})$ and payload $v$ to $\mathsf{stash}$\;
  Let $\rho$ refer to the new stash entry and index it in $\mathcal{I}$ with owner $\mathsf{stash}$\;
}{
  $\mathsf{ans}\gets\bot$\;
  \If{$\mathcal{I}\ne\varnothing$}{
    $\rho\gets\mathcal{I}.\mathsf{deleteMin}()$\;
    Save the priority and payload of $\rho$ as $\mathsf{ans}$\;
    \eIf{$\rho.\mathsf{owner}=\mathsf{stash}$}{Set $b_j\gets0$, where $\rho$ points to $\mathsf{stash}[j]$\;}{
      $r\gets\rho.\mathsf{owner}$\;
      Remove $\rho$ from $\mathcal{C}_r$ and release its storage\;
$c_r\gets c_r+1$\;
    }
  }
  $\widehat{Q}(x)\gets\max\{\widehat{Q}(x)-1,0\}$ for all $x$\;
}
$i\gets i+1$\;
\If{$i=B$ and $T>B$}{$\mathsf{CloseEpoch}()$\;}
\If{$\mathsf{op}=\Extract()$}{\KwRet{$\mathsf{ans}$}\;}
\end{algorithm}


\paragraph{Prefetching.}
Each bin read transfers all $2B$ slots, including padding, into a reusable private buffer. Let $j_r(0)=0$, and $j_r(x)=\min\{j\mid q_{r,j}\ge x\}$ for $x>0$. Then reading through bin $j_r(x)$ covers all records with key at most $x$. Only the last bin may contain larger keys. The frontier $f_r$ never decreases. When $\rho_{r,k}$ is first prefetched from $r$, its index $k$ is obtained by incrementing $\seen_r$. We retain a copy in $\mathcal{C}_r$ exactly when $k>c_r$. Existing counters continue across refreshes.

\begin{algorithm}[!htb]
\SetAlgoCaptionSeparator{.}
\caption{Refresh}
\label{alg:refresh}
\KwIn{The committed segments and the reflected estimate $\widehat{Q}$}
\KwOut{Updated frontiers, local sets, and private index $\mathcal{I}$}
Find the smallest $x\in[V]$ satisfying $\widehat{Q}(x)\ge B+2E$\;
If no such $x$ exists, set $x\gets V$\;
\ForEach{committed segment $r$ in public order}{
  $j^\star\gets\max\{f_r,j_r(x)\}$\;
  \While{$f_r<j^\star$}{
    Read all $2B$ slots of bin $f_r+1$ of $r$ into private buffer $X$\;

    \ForEach{slot $\rho$ of $X$, in physical order}{
      \If{$\neg\rho.\mathsf{pad}$ and $\neg\rho.\mathsf{dead}$}{
        $\seen_r\gets\seen_r+1$\;
        \If{$\seen_r>c_r$}{
          Add a private copy $\rho'$ of $\rho$ to $\mathcal{C}_r$, with $\rho'.\mathsf{owner}=r$\;
          $\mathcal{I}.\mathsf{insert}(\rho')$\;
        }
      }
    }
    $f_r\gets f_r+1$\;
  }
}
\end{algorithm}

\paragraph{Epoch boundaries.}

At an epoch boundary, we create a level-$0$ segment of capacity $B$. The numbers of committed segments at successive levels are the digits of a counter in base $\ell$. Completing an epoch increments this counter. If no carry occurs, the new segment is committed at level $0$. If the resulting carry passes through levels $0,\ldots,j-1$, we merge the level-$0$ segment with all $\ell-1$ segments on each of those levels into a single segment at level $j$. The input segments are called the \emph{sources}, and the output segment is their \emph{parent}. The merge creates only this final parent, with no intermediate segments. The number of sources is $1+j(\ell-1)\le K$, and their total capacity is
\begin{align*}
B+(\ell-1)\sum_{i=0}^{j-1}B\ell^i=B\ell^j.
\end{align*}
Each position in the core operation sequence belongs to the interval of at most one created segment per level. In particular, each insertion contributes to at most $d+1$ summaries.

To build the new level-$0$ segment, we copy all stash records, including extracted ones, into a separate private buffer $X$ and sort it by key. We sort each private buffer by inserting its keys and record pointers into a temporary integer set and repeatedly deleting the minimum. The sorted records of the level-$0$ segment are written into one bin padded to $2B$ slots, leaving $X$ unchanged. The level-$0$ segment descriptor has $m_{0,r}=B$, $m_r=|X|$, directory endpoints $0,V$, an empty local set, and zero counters. The copy of $\mathsf{stash}[i]=(k_i,v_i,b_i)$ retains $(k_i,v_i)$ and has $\mathsf{dead}=1-b_i$ and $\mathsf{pad}=0$. Unused slots have $\mathsf{pad}=1$. The procedure $\mathsf{MergeSegments}$ builds a parent from the sources selected by the public carry. It preserves all original records, transfers their deletion state, and computes a fresh summary. Algorithm~\ref{alg:merge} gives its implementation.

Before the first completed epoch, a private binary heap implements $\mathcal{I}$. The segment hierarchy is empty and the estimates remain zero. At the first epoch boundary, we initialize the integer set and transfer the live entries. If $T\le B$, the entire history remains private and no segment is created. An incomplete final epoch is left in the stash. Parameters and record slots are initialized only when needed. A zero-length history performs no setup.

\begin{algorithm}[!htb]
\SetAlgoCaptionSeparator{.}
\caption{CloseEpoch}
\label{alg:close}
\KwIn{A completed epoch with $i=B$, and public operation bound $T>B$}
\KwOut{The updated segment hierarchy and local sets for the next epoch}
\If{$t_{\mathrm{call}}=B$}{
  Initialize the integer-set constants and transfer live heap entries into $\mathcal{I}$\;
}
\tcp{Construct the replacement segment}
Copy every $\mathsf{stash}[j]$ into private buffer $X$, setting $\mathsf{dead}=1-b_j$ for its copy\;
Sort $X$ by key using a temporary integer set\;
Create a level-$0$ segment $r_0$ from $X$, with its descriptor and counters initialized as above\;
Write $X$ to its single bin, padding to $2B$ slots and leaving $X$ unchanged\;
$\mathsf{BuildSummary}(r_0)$ using the private sorted buffer $X$\;
$\mathcal{R}_{\mathrm{old}}\gets$ committed segments selected by the public carry\;
\eIf{$\mathcal{R}_{\mathrm{old}}=\varnothing$}{$r_{\mathrm{new}}\gets r_0$\;}{
  $m_0\gets B+\sum_{r\in\mathcal{R}_{\mathrm{old}}}m_{0,r}$\;
  $r_{\mathrm{new}}\gets\mathsf{MergeSegments}(\{r_0\}\cup\mathcal{R}_{\mathrm{old}},m_0)$\;
}
\tcp{Replace the summaries and commit}
\ForEach{$r\in\mathcal{R}_{\mathrm{old}}$}{
  \ForEach{entry $(z,\omega)$ streamed from the summary list of $r$}{
    $\widehat{Q}(x)\gets\widehat{Q}(x)-\omega$ for all $x\ge z$\;
  }
}
\ForEach{entry $(z,\omega)$ streamed from the summary list of $r_{\mathrm{new}}$}{
  $\widehat{Q}(x)\gets\widehat{Q}(x)+\omega$ for all $x\ge z$\;
}
$\widehat{Q}(x)\gets\max\{\widehat{Q}(x),0\}$ for all $x$\;
Replace $\mathcal{R}_{\mathrm{old}}$ by $r_{\mathrm{new}}$ in the segment hierarchy\;
\tcp{Retire old copies and prepare the next epoch}
Traverse the local sets of $\mathcal{R}_{\mathrm{old}}$ and the live stash entries, deleting their entries from $\mathcal{I}$\;
Release those local sets and clear $\mathsf{stash}$, releasing its slots\;
$i\gets0$\;
$\mathsf{Refresh}()$\;
\end{algorithm}

During maintenance, the existing segment hierarchy, source counters, and stash originals remain available until the new segment and its summary are complete. Merge scans use separate counters and buffers. Commit replaces $\mathcal{R}_{\mathrm{old}}$ by $r_{\mathrm{new}}$ in the hierarchy and updates $\widehat{Q}$ in one event. It then removes the retired entries from $\mathcal{I}$ by traversing $\mathcal{C}_r$ for $r\in\mathcal{R}_{\mathrm{old}}$ and the live stash entries. It does not scan unused private storage. Unmerged segments keep their state. A new segment starts with $\mathcal{C}_r=\varnothing$ and zero counters, followed by a refresh. The current answer remains in its output register until the call returns. All simultaneous storage is counted in the space bound.


\subsection{Merging and Summary Construction}
\label{sec:segments}

\paragraph{Merging segments.}

The source summaries determine bin boundaries with at most $2B$ records per output bin. A merge reads each source in order and retains at most one unfinished bin per source. Each output bin fits in private memory and can be sorted in linear work by Lemma~\ref{lem:small-integer-sets}. We first give this merge using the summary interface, then implement the interface.

A level-$0$ segment has one bin with boundaries $0,V$. To create a parent, let $\mathcal{R}$ be the ordered collection of source segments, including the new level-$0$ segment. Write $m_r$ and $m_{0,r}$ for source $r$'s original-record count and public capacity, respectively. The parent $P$ has original-record count $m=\sum_{r\in\mathcal{R}}m_r$ and public capacity $m_0=\sum_{r\in\mathcal{R}}m_{0,r}$, with $m\le m_0$; the count $m$ includes records already extracted. Let $S(x)=\sum_{r\in\mathcal{R}}S_r(x)$. This sum is a function of released information. Let $F_P(x)$ count the parent's original records with key at most $x$. The summary interface gives $S(V)=m$ and $\|S-F_P\|_\infty\le KZ$. For $1\le k<m_0/B$, let $q_{P,0}=0$, $q_{P,m_0/B}=V$, and
\begin{align*}
q_{P,k}=
\begin{cases}
\min\{x\in[V]\mid S(x)\ge kB\},& kB\le m,\\
V,&kB>m.
\end{cases}
\end{align*}

\begin{lemma}
\label{lem:bin-capacity}
Every parent bin contains at most $B+2KZ+1\le 2B$ original records.
\end{lemma}

\begin{proof}
Whenever $q_{P,k}$ is chosen as a threshold, $S(q_{P,k})\ge kB$, while the preceding integer has summary value less than $kB$. Uniform error and distinct keys give
\begin{align*}
kB-KZ\le F_P(q_{P,k})\le kB+KZ+1.
\end{align*}
The difference between the upper bound at $q_{P,k}$ and the lower bound at $q_{P,k-1}$ is at most $B+2KZ+1$. For the first interval, the same bound follows from $F_P(0)=0$. For the final nonempty interval, $F_P(V)=m$ and either $m<kB$ or $k=m_0/B$ with $m\le m_0$, so the preceding lower bound again gives the claimed capacity. Bins between repeated boundaries at $V$ contain no records. The condition $B\ge\ell^{12}+4KZ+4$ gives $B+2KZ+1\le 2B$.
\end{proof}

Algorithm~\ref{alg:merge} constructs the data and directory before generating the new summary. Its source-specific cursors refer only to records already read. In particular, records beyond the current parent boundary remain in their source buffers. The local counter $p_r$ counts bins read during this merge. The counter $o_r$ counts encountered records that were live when the source was created. They are separate from the stored prefetch and extraction counters.

The parent $P$ reserves fresh, nonoverlapping server regions for its data, directory, summary, and work arrays. Its descriptor stores capacity $m_0$, $m_0/B$ bins, an empty local set, and zero counters. Output bins $P.\mathsf{bin}$ and the bin directory $(q_{P,k})_{k=0}^{m_0/B}$ reside on the server; assignments to these arrays are server writes. The current boundary $q_{P,k}$ is written once and retained in a private register while bin $k$ is being constructed. The source remainders $\mathcal{B}_r$, current input bin $X$, output buffer $\mathcal{O}$, and sorting set are private. Appending a record from $X$ to $\mathcal{O}$ or $\mathcal{B}_r$ stores a copy that remains valid when $X$ is reused. Source segments are traversed in a fixed public order.

\begin{algorithm}[!htb]
\SetAlgoCaptionSeparator{.}
\caption{MergeSegments$(\mathcal{R},m_0)$}
\label{alg:merge}
\KwIn{Ordered sources $\mathcal{R}$, with counts $c_r$ and total capacity $m_0$}
\KwOut{A complete, uncommitted parent segment $P$}
Reserve fresh server regions for the data, directory, summary, and work arrays of $P$\;
$m\gets\sum_{r\in\mathcal{R}}m_r$\;
Initialize the descriptor of $P$ as above, with record count $m_P=m$\;
$q_{P,0}\gets0$\;
\ForEach{$r\in\mathcal{R}$}{
  $\mathcal{B}_r\gets\varnothing,\quad p_r,o_r\gets0$\;
}
\For{$k\gets1$ \KwTo $m_0/B$}{
  \eIf{$k=m_0/B$ or $kB>m$}{$q_{P,k}\gets V$\;}{
    $q_{P,k}\gets\min\{x\in[V]\mid \sum_{r\in\mathcal{R}}S_r(x)\ge kB\}$\;
  }
$\mathcal{O}\gets\varnothing$\;
  \ForEach{$r\in\mathcal{R}$ in its fixed order}{
    Move the prefix of $\mathcal{B}_r$ with keys at most $q_{P,k}$ to $\mathcal{O}$\;
    $j^\star\gets j_r(q_{P,k})$\;
    \While{$p_r<j^\star$}{
      $p_r\gets p_r+1$\;
      Read all $2B$ slots of bin $p_r$ of $r$ into private buffer $X$\;
      \ForEach{slot $\rho$ of $X$, in physical order}{
        \If{$\neg\rho.\mathsf{pad}$}{
          \tcp{Account for extractions since the source segment was created}
          \If{$\neg\rho.\mathsf{dead}$}{
            $o_r\gets o_r+1$\;
$\rho.\mathsf{dead}\gets(o_r\le c_r)$\;
          }
          \eIf{$\rho.\mathsf{key}\le q_{P,k}$}{$\mathcal{O}.\mathsf{append}(\rho)$\;}{
            $\mathcal{B}_r.\mathsf{append}(\rho)$\;
          }
        }
      }
    }
  }
  Sort $\mathcal{O}$ by key using a temporary integer set\;

  Pad $\mathcal{O}$ to $2B$ slots and write the entire bin to $P.\mathsf{bin}[k]$\;
}
$\mathsf{BuildSummary}(P)$\;
\KwRet{$P$}\;
\end{algorithm}

The following lemma gives the merge bound needed for the amortized analysis. Together with Lemma~\ref{lem:streaming-summary}, it shows that constructing a segment takes expected work linear in its capacity.

\begin{lemma}[Correctness and cost of merging]
\label{lem:merge}
Fix the parameters of Section~\ref{sec:layout}. Let $\mathcal R$ contain at most $K$ source segments satisfying the representation and summary guarantees stated there. Assume that their epoch intervals are disjoint and that they are traversed in a fixed public order. Write $m_0=\sum_{r\in\mathcal R}m_{0,r}$ for their total capacity. For each source $r$, assume that the records extracted since its creation are precisely $\rho_{r,1},\ldots,\rho_{r,c_r}$.

The merge phase of Algorithm~\ref{alg:merge}, before the call to $\mathsf{BuildSummary}(P)$, has the following properties.
\begin{enumerate}[label=(\roman*),leftmargin=*]
\item The parent stores every original record exactly once in the bin prescribed by its key. Records within each bin are sorted by key. Each record's deletion bit is $1$ if and only if it was extracted before the parent was created.
\item The phase uses $O(m_0)$ word operations and probes and $O(KB)$ additional private words on every execution.
\item Its server access pattern is determined solely by the source summaries, source bin directories, and public parameters.
\end{enumerate}
\end{lemma}

\begin{proof}
For (i), we maintain the invariant that, after writing parent bin $k$, every original record with key at most $q_{P,k}$ has been output exactly once. To construct this bin, the algorithm consumes the appropriate buffered prefixes and reads each source through bin $j_r(q_{P,k})$. Records with larger keys remain in $\mathcal B_r$, and a buffered record is removed when transferred to $\mathcal O$. Thus $\mathcal O$ contains exactly the records with keys in $(q_{P,k-1},q_{P,k}]$. Sorting $\mathcal O$ gives the required order, and the final boundary $V$ ensures that every original record is included.

Records marked dead in a source retain that bit. For the other records, the scan increments $o_r$ in the order $\rho_{r,1},\rho_{r,2},\ldots$. When $\rho_{r,i}$ is processed, $o_r=i$, so its deletion bit is set exactly when $i\le c_r$. The assumed deletion prefix therefore gives the correct bit for every record. These bits remain attached to buffered records, and the source counters are unchanged.

For (ii), each source buffer $\mathcal B_r$ contains only records from the last bin read from that source. Earlier bins lie entirely below the current output boundary. Each source buffer therefore occupies $O(B)$ private words. Lemma~\ref{lem:bin-capacity} bounds $\mathcal O$ by $2B$ records. Together with the current input bin and the sorting workspace, the additional private space is $O(KB)$.

Each physical input slot is inspected at most once, and each buffered record is transferred to the output once. There is at most one unsuccessful prefix test per source per output bin. Scanning and buffer processing therefore cost $O(m_0+Km_0/B)=O(m_0)$ word operations. By Lemma~\ref{lem:small-integer-sets}, sorting each output bin costs $O(B)$ word operations because $B=w^{O(1)}$. There are $m_0/B$ output bins, so sorting, padding, and writing them cost $O(m_0)$ in total.

Computing a parent boundary by binary search over the key universe and the source summary lists costs $O(K\log^2 T)$ word operations and probes. Locating that boundary in the source directories costs $O(K\log T)$. Summed over all output bins, these searches cost $O\!\left(m_0K\log^2 T/B\right)=O(m_0)$ under the stated parameter choices. This proves the work and probe bounds on every execution.

For (iii), the source summaries determine all parent boundaries and the search paths used to find them. The source directories then determine which bins are read. Sources are processed in a fixed public order, each input bin is read in full, and each output bin is written to its prescribed location using exactly $2B$ slots. Directory entries are also written to prescribed locations. Keys and deletion bits affect only private computation and the contents written, so they do not otherwise affect the server access pattern.
\end{proof}

The parent preserves all original records, and $\mathsf{BuildSummary}$ recomputes their dyadic counts rather than using the source estimates as counts. Its summary therefore has error $Z$, independently of the error $KZ$ used to choose its bins. Accuracy does not deteriorate across levels. Successive fresh releases instead contribute to the privacy composition in Section~\ref{sec:joint-trace}.

\paragraph{Private summaries.}

We construct a summary satisfying the interface in Section~\ref{sec:layout}, using the public parameters specified in Section~\ref{sec:layout}. The mechanism uses the complete dyadic tree over $[V]$, with $h$ levels. The final summary must approximate every prefix of original insertions, independently of deletion state. The scale $L$ gives each geometric draw success probability $1/L$ and satisfies $L\ge4h(d+1)/\eta$. The parameter $R_0$ controls the truncation probability, and $H$ is the truncation threshold for each geometric draw.

Consider a segment $r$ containing $m$ original records, with exact cumulative count $F_r(x)$. The number $m$ is public because operation types are public. Let $c_u$ count original records in dyadic node $u$, including records already extracted. Replacing one original record changes at most $2h$ coordinates of the complete count vector, each by at most one. This sensitivity concerns original fields and does not depend on which records have been extracted.

Set $q=1-1/L$. Conceptually, for every dyadic node independently draw two zero-based geometric variables $G_{u,1},G_{u,2}$ with success probability $1/L$, and define
\begin{align*}
\xi_u=\min\{G_{u,1},H\}-\min\{G_{u,2},H\},\quad y_u=c_u+\xi_u.
\end{align*}
Retain the pair $(u,y_u)$ exactly when $y_u\ge2H+1$. Denote the resulting sparse vector, including node identifiers and noisy counts, by $\mathcal{S}_r$. The privacy analysis concerns this sparse vector. The final summary list below is obtained by postprocessing it. The conceptual full vector is never stored by the implementation.

The operation $\mathsf{CappedGeom}(L,H)$ returns $\min\{G,H\}$ for a zero-based geometric variable $G$ of success probability $1/L$. This draw assigns mass $(1-q)q^j$ to $0\le j<H$ and mass $q^H$ to $H$. It can be sampled exactly using constant private space, at most $H$ Bernoulli trials on every execution, and at most $L$ trials in expectation. Sampling makes no server accesses. Appendix~\ref{app:word-ram} gives the implementation using independent uniform random words.

\begin{lemma}[Privacy of the noisy dyadic counts]
\label{lem:bounded-noise}
Fix the public parameters of Section~\ref{sec:layout} and a segment $r$ containing a public number $m$ of original records. Let $u$ range over the nodes of the complete dyadic count tree on $[V]$, which has $h$ levels. For each node $u$, let $c_u$ count the original records of $r$ whose keys lie in its interval, including records already extracted. Set $y_u=c_u+\xi_u$, where $\xi_u$ is the difference of two independent geometric samples with success probability $1/L$, each capped at $H$, as defined above. The noise variables $\xi_u$ are independent across nodes.

Let $(\eta,\delta)$ be the privacy parameters of the core, and let $d+1$ be the number of levels in the segment hierarchy. The mechanism that releases the complete vector $(y_u)_u$ is $(\eta/(d+1),\delta/(d+1))$-differentially private under replacement of one original record in $r$, with $m$ fixed. The retained vector $\mathcal{S}_r=\{(u,y_u)\mid y_u\ge2H+1\}$ satisfies the same privacy guarantee by postprocessing.
\end{lemma}

\begin{proof}
Fix two neighboring multisets of $m$ original records, and let $c_u$ and $c'_u$ be their respective counts at node $u$. Set $\mathcal D=\{u\mid c_u\ne c'_u\}$ and $k=|\mathcal D|$. Replacing one record changes at most two node counts per level, each by at most one, so $k\le2h$. Let $q=1-1/L$. The difference of two independent uncapped geometric variables has probability mass proportional to $q^{|z|}$. On the changed coordinates, independent uncapped difference noises therefore give a likelihood ratio at most
\begin{align*}
q^{-k}\le\exp(2k/L)\le\exp(4h/L)\le e^{\eta/(d+1)}.
\end{align*}
Here we use $L\ge2$ and $L\ge4h(d+1)/\eta$.

For each of the two neighboring inputs, define a hybrid output distribution by using uncapped difference noise at coordinates in $\mathcal D$ and bounded noise elsewhere. The unchanged coordinates have identical independent distributions under the two inputs. The two hybrid output distributions therefore satisfy the preceding pure likelihood-ratio bound.

Under the coupling using the same geometric variable, a one-sided capped draw and its uncapped counterpart differ only if $G>H$, an event of probability $q^{H+1}$. For each input, its bounded-noise output distribution is consequently within total variation distance $2kq^{H+1}$ of its hybrid output distribution. Combining these two total variation bounds with the likelihood-ratio bound between the hybrid distributions gives additive slack at most
\begin{align*}
2k(1+e^{\eta/(d+1)})q^{H+1}
&\le4h\kappa e^{-R_0}
\le4h\kappa2^{-R_0}
\le\frac{\delta}{d+1}.
\end{align*}
The first inequality uses $k\le2h$, $1+e^{\eta/(d+1)}\le1+e^\eta\le\kappa$, and $H=LR_0$, which gives $q^{H+1}\le e^{-(H+1)/L}\le e^{-R_0}$. The last inequality follows from the parameter choice $2^{R_0}\ge4\kappa h(d+1)/\delta$. Thresholding the bounded vector to form $\mathcal{S}_r$ is postprocessing. Only changed coordinates enter the comparison, so no factor for the total number of dyadic nodes appears.
\end{proof}

Because $|\xi_u|\le H$ on every execution, a node with $c_u\le H$ can never be retained. The actual algorithm therefore enumerates and samples only nodes with $c_u\ge H+1$. Its retained vector has exactly the conceptual mechanism's distribution. We call these candidate nodes heavy. At each tree level their intervals are disjoint, so their total number is at most $mh/(H+1)$. Their expected total sampling work is consequently
\begin{align*}
O\!\left(\frac{mhL}{H+1}\right)=O\!\left(\frac{mh}{R_0}\right)=O(m).
\end{align*}
The heavy-node set and the number of rejected candidates remain private. The implementation below hides the candidates that are not retained by writing a fixed number of padded slots.

For any node, set its estimated count to $y_u$ if $(u,y_u)\in\mathcal{S}_r$, and to zero otherwise. For $x>0$, sum these estimated counts over the at most $h$ nodes in the canonical dyadic decomposition of $[1,x]$, obtaining $\widetilde{F}(x)$. Define a monotone function by clipping and taking prefix maxima, with the public total enforced at the final coordinate.
\begin{align*}
G(x)&=\max_{0<y\le x}\min\{m,\max\{0,\widetilde{F}(y)\}\}\qquad(0<x<V),\\
G(0)&=0,\qquad G(V)=m.
\end{align*}

We quantize $G$ in steps of $s$. The bound $Z$ in~\eqref{eq:summary-error} includes the quantization error. For $j=1,\ldots,\lfloor m/s\rfloor$, output the first coordinate $z_{r,j}$ with $G(z_{r,j})\ge js$, with weight $s$. Output one additional pair $(V,m-s\lfloor m/s\rfloor)$ even when its weight is zero. The list length is fixed by $m$, and positions may repeat. The weighted cumulative count represented by this list is $S_r$, as in~\eqref{eq:stored-summary}.

\begin{lemma}[Accuracy of the segment summary]
\label{lem:static-summary}
On every execution, the summary function $S_r$ is nondecreasing, satisfies $S_r(V)=m$, and obeys $\|S_r-F_r\|_\infty\le Z$.
\end{lemma}

\begin{proof}
A retained node has count error at most $H$. An absent node satisfies $c_u+\xi_u\le2H$, hence $c_u\le3H$. The zero estimate for such a node has error at most $3H$. Summing over the at most $h$ canonical nodes of any prefix gives, for every noise outcome,
\begin{align*}
|\widetilde{F}(x)-F_r(x)|\le3hH.
\end{align*}
Clipping to $[0,m]$ and taking prefix maxima do not increase uniform error from the monotone function $F_r$. Setting $G(V)=m$ preserves this bound and monotonicity. For $0<x<V$, the quantized count is $s\lfloor G(x)/s\rfloor$, while $S_r(V)=m$. Quantization loses less than $s$, yielding error at most $s+3hH\le Z$. 
\end{proof}

A merge combines at most $K$ summaries, so its aggregate error is at most $KZ$. By the parameter choices in~\eqref{eq:buffer-parameters} and~\eqref{eq:prefetch-parameters}, we have $B+2KZ+1\le 2B$ and $E<5B/4$.
The first inequality proves the capacity condition used by the merge. The term $\ell^{12}$ amortizes the searches and metadata work, as shown in Section~\ref{sec:probe-accounting}.

\paragraph{Computing the summary.}

The procedure $\mathsf{BuildSummary}(r)$ scans every original record of segment $r$ in sorted bin order and ignores deletion bits. It enumerates only dyadic nodes with count at least $H+1$. A compressed trie skips all other nodes and uses an $O(h)$ private stack. After processing each physical bin $i$, including a bin containing only padding, the scan reaches \emph{checkpoint} $i$, associated with the key interval $(q_{r,i-1},q_{r,i}]$. We add zero-weight points at bin boundaries to close all nodes whose right endpoints lie in the completed bin. Appendix~\ref{app:enumeration} gives the iterator.

\begin{lemma}[Heavy-node enumeration]
\label{lem:heavy-iterator}
Let $r$ be a segment of capacity $m_0$ containing $m$ original records stored in key order in its bins. A node of the dyadic tree over $[V]$ is heavy if its interval contains at least $H+1$ original records. There exists a streaming iterator that privately enumerates every heavy node exactly once, together with its exact count. A node whose right endpoint lies in $(q_{r,i-1},q_{r,i}]$ is assigned to checkpoint $i$ and enumerated by the end of bin $i$. Within each checkpoint, outputs are ordered by right endpoint, with ties broken by increasing height. Excluding noise sampling, the iterator uses
\begin{align*}
O\!\left(m_0+\frac{mh}{H+1}\right)
\end{align*}
word operations and $O(h)$ additional private words beyond the current bin buffer, where $h$ is the number of levels in the dyadic tree.
\end{lemma}

There are at most $mh/(H+1)$ heavy nodes. Lemma~\ref{lem:heavy-iterator} charges trie traversal to input records, bin boundaries, and emitted nodes.

At each bin boundary, the procedure writes exactly
\begin{align*}
R_{\mathrm{slot}}=h\left(1+\left\lfloor\frac{2B}{H+1}\right\rfloor\right)
\end{align*}
slots at prescribed server addresses. It buffers retained nodes in order of right endpoint and then height, and fills unused slots with padding. Each slot contains a validity bit, a node identifier, and a noisy count. At each height, at most one node ending in a bin crosses its left boundary. The remaining heavy intervals lie inside the bin and are disjoint. Their number is at most $\lfloor 2B/(H+1)\rfloor$. Thus the retained nodes at each checkpoint fit in the buffer of $R_{\mathrm{slot}}$ slots.

The reads and the number and addresses of these writes are fixed by the old layout. A retained node is assigned to the first bin boundary at or beyond its right endpoint. Its position is determined by the order of retained nodes. Thus the padded array is postprocessing of $\mathcal{S}_r$ and the old layout. Rejected candidates and local sampling times cause no further server accesses. The array contains
\begin{align*}
P_0=\frac{m_0}{B}R_{\mathrm{slot}}=O\!\left(\frac{m_0h}{H+1}+\frac{m_0h}{B}\right)
\end{align*}
slots and remains on the server.

To compute $\widetilde{F}$ by a coordinate sweep, we use \emph{endpoint events} $(z,\Delta)$, where $z\in[V+1]$ is a sweep coordinate and $\Delta$ is the signed increment to the running estimate at $z$. The coordinate $V+1$ is a sentinel beyond the key universe.

Read the sparse slots in order. For each valid root count $y$, write events $(V,y)$ and $(V+1,-y)$. For a valid left child with interval $[l_u,r_u]$ and noisy count $y$, write
\begin{align*}
(r_u,y),\qquad (r_u+(r_u-l_u+1),-y).
\end{align*}
For a right child or a padding slot, write two zero events at $V+1$. Also write $(V,0)$. Each sparse slot therefore causes exactly two event writes. A left child contributes to the canonical prefix decomposition precisely between its two event coordinates. The root contributes only at $V$.

Sort the event array on the server by coordinate using in-place heapsort with constant workspace. The sort is postprocessing of the padded sparse array. Sum all deltas at a coordinate before updating the prefix maximum. The resulting sweep computes $G$, with $G(V)=m$, and emits a weighted position whenever $G$ reaches the next multiple of $s$. Advance this threshold by addition and write the final residual entry at $V$, even when its weight is zero. Both the events and the summary remain on the server. This procedure realizes the cumulative summary in Lemma~\ref{lem:static-summary} without storing either array privately.

\begin{lemma}[Summary construction]
\label{lem:streaming-summary}
Fix the parameters of Section~\ref{sec:layout}. Let $r$ be a segment of public capacity $m_0\ge B$, with its original records already stored in key order in the prescribed bins and its bin directory constructed.

The procedure $\mathsf{BuildSummary}(r)$ constructs the retained vector $\mathcal{S}_r$ and a weighted list representing $S_r$, with the privacy and accuracy guarantees of Lemmas~\ref{lem:bounded-noise} and~\ref{lem:static-summary}. Both outputs are stored on the server. It uses $O(m_0)$ expected total work. On every execution, it incurs $O(m_0)$ probes and uses $O(B+h+R_{\mathrm{slot}})$ additional private words.

Its server access pattern is determined solely by the existing bin directory and storage addresses, the generated vector $\mathcal{S}_r$, and public parameters.
\end{lemma}

\begin{proof}
Let $m\le m_0$ be the number of original records in $r$. A node with $c_u\le H$ has $y_u\le2H$ and can never be retained. Sampling only heavy nodes therefore produces exactly the distribution of $\mathcal{S}_r$ in Lemma~\ref{lem:bounded-noise}. The subsequent sweep and quantization construct the weighted list representing $S_r$ as in Lemma~\ref{lem:static-summary}. This list is a deterministic function of $\mathcal{S}_r$ and public parameters, which establishes the stated privacy and accuracy guarantees.

Let $W_{\mathrm{summary}}$ denote the total work of this procedure. Lemma~\ref{lem:heavy-iterator} bounds the input scan and enumeration by $O(m_0+mh/(H+1))$ word operations. Sampling each heavy node takes $O(L)$ expected word operations. Generating the $P_0$ padded slots costs $O(P_0)$ work and probes. Sorting the event array costs $O(P_0\log T)$ work and probes, and the subsequent sweep and quantization cost $O(P_0+m/s+1)$. All stored quantities fit in a constant number of words. Therefore
\begin{align*}
\E[W_{\mathrm{summary}}]
&=O\!\left(m_0+\frac{mhL}{H+1}+\frac{m_0h\log T}{H+1}+\frac{m_0h\log T}{B}+\frac ms+1\right)\\
&=O(m_0),
\end{align*}
using $m\le m_0$, $H=LR_0$, $R_0\ge\ell^2$, $h=O(\ell)$, and $B=\Omega(\ell^{12})$. The input scan, padded writes, event sort, and output sweep use $O(m_0)$ probes for every outcome of the samples, since sampling makes no server accesses.

The private workspace consists of the current bin buffer, the iterator stack, at most $R_{\mathrm{slot}}$ buffered tuples, and a constant number of sampling and sweep registers. These occupy $O(B+h+R_{\mathrm{slot}})$ words on every execution. The padded array, event arrays, and summary list remain on the server.

The input scan and the addresses of the $R_{\mathrm{slot}}$ writes at each checkpoint are fixed by the existing bin directory, storage addresses, and public parameters. The retained pairs are assigned to checkpoints by their right endpoints and ordered within each checkpoint by endpoint and height. Thus the padded array is determined by $\mathcal{S}_r$ and the bin directory. All subsequent event generation, sorting, and sweeping are deterministic operations on this array and public parameters. The complete server access pattern therefore has the stated dependence.
\end{proof}

To construct a level-$0$ segment, we copy the stash originals and their current live bits into a separate $O(B)$ private output buffer, sort that buffer, and use the same iterator with one bin and no server input reads. The original stash slots and their live-index pointers remain unchanged until commit. This copy is covered by the $O(B)$ output and sorting workspace. We still write the prescribed padded sparse slots before server-side event processing.

Every segment's data, directory, sparse-event, and summary locations are initialized before use, and these writes are included in the bounds above. The prescribed checkpoint and residual slots are written even for an empty segment. Reserved address ranges are initialized only as needed.

\subsection{Hiding Operation Types}
\label{sec:wrapper}

We use the core as a target priority queue to implement a source priority queue with private operation types. Fix the source operation bound $T$ and priority domain $[U]$, where $U=T^{O(1)}$. Here $T$ bounds external operations, and the target is instantiated with operation bound $3T$.

The wrapper's dummy records are ordinary inserted records of the target queue and contribute to its counts and summaries. They are distinct from the padding slots used in bins and summary work arrays. Every source round $t\in[T]$ invokes the public target pattern $\Insert,\Insert,\Extract$. On source operation $\Insert(p,v)$, the wrapper inserts a real record with priority $T+p$ and tagged payload $(\mathsf{real},v)$, followed by a guard with priority $t$ and tagged payload $(\mathsf{guard},0)$. It then performs an extraction and discards the answer. On a source $\Extract$, the wrapper inserts two records with priorities $T+U+2t$ and $T+U+2t+1$, both tagged $\mathsf{dummy}$, and performs the extraction. A real answer is returned after subtracting $T$ from its priority and removing its tag. A dummy answer is translated to $\bot$.

Lemma~\ref{lem:comparison-hide-types} establishes the wrapper's correctness, privacy, and overhead bounds.

\section{Analysis}
\label{sec:analysis}

We establish correctness, privacy, and resource bounds for the construction of Section~\ref{sec:construction}, proving Theorem~\ref{thm:main}. We then extend the bounds to varying privacy parameters.

\subsection{Correctness}
\label{sec:correctness}

\paragraph{Reflecting insertion counts.}

Fix a threshold $x$. Let $A_t(x)$ count all original records inserted by event $t$ with key at most $x$, including extracted records, and let $Q_t(x)$ count the currently live records in that prefix. The counter $N_t$ counts extraction calls as in Section~\ref{sec:operations}. All counts start at zero.

We use the event index of Section~\ref{sec:operations}, so an extraction and a subsequent replacement of committed segments are separate events. An auxiliary event changes neither $A$ nor $N$.

\begin{lemma}[Reflection and stability]
\label{lem:reflection}
Define $X_t(x)=A_t(x)-N_t$. Then
\begin{align}
Q_t(x)=X_t(x)-\min_{0\le u\le t}X_u(x).
\label{eq:reflection}
\end{align}
If an estimated process satisfies $|\widehat{A}_t(x)-A_t(x)|\le E$ at every real and auxiliary event, define $\widehat{X}_t(x)=\widehat{A}_t(x)-N_t$ and $\widehat{Q}_t(x)=\widehat{X}_t(x)-\min_{u\le t}\widehat{X}_u(x)$. Then $|\widehat{Q}_t(x)-Q_t(x)|\le2E$ for all $t,x$.
\end{lemma}

\begin{proof}
An insertion of a record with encoded key $k$ adds $\one_{k\le x}$ to $Q(x)$. An extraction applies $Q(x)\leftarrow[Q(x)-1]_+$. Whenever $Q(x)>0$, the global minimum is also at most $x$. An auxiliary event has zero true increment. Thus $Q_t=[Q_{t-1}+X_t-X_{t-1}]_+$, whose solution from zero is~\eqref{eq:reflection}. If $\widehat{X}$ is within $E$ of $X$, their historical minima are also within $E$. Subtracting gives the claimed $2E$ bound.
\end{proof}

The same running-minimum identity solves the estimated recurrence in Section~\ref{sec:operations}, including negative increments caused by summary replacements. Thus the maintained $\widehat{Q}$ is the reflected process in Lemma~\ref{lem:reflection}.

\paragraph{Applying the estimate.}
Every original insertion outside the stash belongs to one committed segment, and the stash contains at most $B$ records. Lemma~\ref{lem:static-summary} therefore gives, at every real or auxiliary event,
\begin{align*}
|\widehat{A}(x)-A(x)|\le KZ+B=E.
\end{align*}
This holds both before and after a replacement of committed segments. Lemma~\ref{lem:reflection} consequently gives live-prefix error at most $2E$.

\begin{lemma}[A cutoff for exact prefetching]
\label{lem:cutoff}
Suppose $\|\widehat{Q}-Q\|_\infty\le2E$. If $x$ is the smallest integer satisfying $\widehat{Q}(x)\ge B+2E$, then
\begin{align*}
B\le Q(x)\le B+4E+1.
\end{align*}
If no such $x$ exists, the cutoff is $V$ and the total number of live records is less than $B+4E$.
\end{lemma}

\begin{proof}
The lower bound follows from the estimation error. If $x>1$, minimality gives $\widehat{Q}(x-1)<B+2E$, so $Q(x-1)<B+4E$. Distinct composite integer keys imply $Q(x)-Q(x-1)\le1$. The case $x=1$ uses $Q(0)=0$. If there is no crossing, apply the error bound at $V$.
\end{proof}

\paragraph{Segment invariant.}
After each core operation, the following invariant holds for every $\rho_{r,k}$. It has been extracted if and only if $k\le c_r$, has been prefetched from $r$ if and only if $k\le\seen_r$, and belongs to $\mathcal{C}_r$ if and only if $c_r<k\le\seen_r$. In particular, $c_r\le\seen_r$. The correctness proof below establishes this invariant together with exactness.

The prefix through bin $j_r(x)$ covers every record with key at most $x$, and only its last bin may contain larger keys. The frontier may extend beyond $j_r(x)$ because of an earlier refresh. Since $f_r$ only advances, $\seen_r$ numbers records live at segment creation across all refreshes.

\begin{lemma}[Local sets and exact online answers]
\label{lem:local-set}
After each refresh, $\bigcup_r\mathcal{C}_r$ contains the $B$ live records with the smallest keys, or all live records if fewer remain. Every segment $r$ satisfies
\begin{align*}
|\mathcal{C}_r|\le 3B+4E+1=O(B).
\end{align*}
These sets and the next epoch's stash suffice to answer all its operations exactly.
\end{lemma}

\begin{proof}
Coverage follows from Lemma~\ref{lem:cutoff} and the complete-prefix reads. For the size bound, consider the last refresh that advanced a particular segment's frontier, and let $x$ be the cutoff used at that refresh. The segment's entire read prefix then ended at bin $j_r(x)$. At that refresh, at most $B+4E+1$ live records globally had keys at most $x$, and the segment's boundary bin contained at most $2B$ additional original records. The segment receives no new insertions, so $\mathcal{C}_r$ can only shrink until its frontier next advances or the segment is merged.

During the next epoch, each new insertion is retained in the stash. Each extraction selects the minimum among the local sets and live stash records and updates the corresponding $c_r$ or stash live bit. At most $B$ extractions occur. A newly inserted small record can delay the extraction of old records but cannot cause the $(B+1)$st smallest old record to precede the first $B$. Thus the retained old prefix is sufficient, including when the queue becomes empty.
\end{proof}

\begin{lemma}[Exactness and termination]
\label{lem:correctness}
On every random execution, the core returns exact answers and terminates. At each commit, the new segment has the correct deletion bits and the segment hierarchy together with the stash represents exactly the queue's state.
\end{lemma}

\begin{proof}
Initially all records are in the stash, so the private heap answers exactly. Suppose the segment invariant and the coverage property hold at the start of an epoch. Lemma~\ref{lem:local-set} supplies the old records needed for that epoch, and every new insertion is indexed privately. Each extraction therefore selects the true minimum. If it removes a record from segment $r$, that record is the next in the segment's creation-time live order. Removing it from $\mathcal{C}_r$ and incrementing $c_r$ preserve the segment invariant. The ordinal rule of Algorithm~\ref{alg:merge} consequently transfers exactly the first $c_r$ deletions to the parent.

A newly committed segment starts with an empty local set and $f_r=c_r=\seen_r=0$. Refresh extends each segment's previously read prefix and inserts precisely the records whose ordinals among records live at segment creation exceed $c_r$. Thus refresh preserves the segment invariant without reintroducing extracted records.

Every output bin fits by Lemma~\ref{lem:bin-capacity}. A new summary is computed from all original records. Lemma~\ref{lem:static-summary} therefore bounds its error for every noise outcome. Commit replaces the summaries in one auxiliary event and preserves the insertion-error bound. Lemma~\ref{lem:reflection} then gives live-prefix error at most $2E$. Lemmas~\ref{lem:cutoff} and~\ref{lem:local-set} imply that the next refresh retains enough old live records to answer the following epoch exactly. Old state is retained until the replacement is complete. Each scan and sort is finite, and each capped geometric sample uses at most $H$ trials.
\end{proof}

\subsection{Privacy Analysis}
\label{sec:joint-trace}

We use postprocessing and sequential composition for differentially private releases. Releases with budgets $(\eps_i,\delta_i)$ compose to budget $(\sum_i\eps_i,\sum_i\delta_i)$ when each release satisfies its privacy guarantee conditional on the preceding releases.

For the core, we restrict $\neigh$ to pairs in which the differing operation is an insertion in both inputs, and denote this relation by $\neigh_{\rm ins}$. The operation-type sequence is treated as public.

\begin{lemma}[Privacy of the core]
\label{lem:core-privacy}
For every fixed public type history, the core's complete access pattern is $(\eta,\delta)$-differentially private under $\neigh_{\rm ins}$.
\end{lemma}

\begin{proof}
Fix the public type history. For every segment $r$ on the public creation schedule, let $\mathcal{S}_r$ be its retained-node vector from Lemma~\ref{lem:bounded-noise}. This vector specifies which dyadic nodes pass the noisy threshold, together with their endpoints and noisy counts. It is generated from all original insertions in the segment's fixed time interval, including records already extracted, using independent fresh samples. The final summary list is a postprocessing of this vector. The analysis uses these vectors because the sparse-event sort may reveal information not determined by the final summary lists.

If two histories differ in one insertion, only the level-$0$ segment and subsequent ancestor segments containing that insertion have different summary-generation inputs. There are at most $d+1$ such segments. Conditional on any earlier released vectors, an affected segment still applies a fresh $\bigl(\frac{\eta}{d+1},\frac{\delta}{d+1}\bigr)$-DP mechanism to its fixed original-record input. A segment whose input is unchanged contributes no additional privacy loss. Sequential composition therefore makes the complete retained-vector collection $(\eta,\delta)$-DP. Physical cuts may depend on earlier vectors, but neither the membership of a segment nor the distribution of its new retained vector depends on those cuts. Payload replacements leave every summary-generation input unchanged.

The entire server trace is a postprocessing of this collection and the public type history. Merge membership and source order are public. Earlier vectors determine source summaries, parent boundaries, directories, and the physical prefixes read by a merge. During a scan that generates a new summary, each physical bin is read in full and each checkpoint writes exactly $R_{\mathrm{slot}}$ tuples to predetermined positions. The valid tuples in this group are precisely the retained nodes whose right endpoints belong to that checkpoint's interval, sorted in canonical order and padded to the fixed slot count. Their group and order are determined by the retained vector and the previously released cuts. Exact counts, rejected heavy nodes, and local discovery times are not released.

Expanding these fixed arrays into endpoint events, sorting the events, grouping equal coordinates, and producing weighted summary entries $(z_{r,i},\omega_{r,i})$ all use only the released vectors and public parameters. Metadata updates, cutoff searches, and later physical prefetch frontiers are functions of the same information. The records selected for extraction, deletion bits, ordinals among records live at segment creation, and actual record counts within bins affect private computation and server contents. They do not select server addresses. Internal instructions and their elapsed time are outside the trace model. The same information also determines which call makes each probe. Extraction updates follow the public type history, and epoch maintenance is completed within the closing call fixed by the public call counter. The sequence of call traces, including empty traces, is therefore $(\eta,\delta)$-differentially private by postprocessing.

\end{proof}

\paragraph{Hiding operation types.}
We now analyze the wrapper from Section~\ref{sec:wrapper}, which gives each external call the same public internal type pattern.

\begin{lemma}[Hiding operation types by fixed padding]
\label{lem:comparison-hide-types}
Suppose a target priority queue is exact and, for every fixed public type sequence, its access pattern, grouped by target call, protects replacement of one insertion's private fields with $(\eta,\delta)$ differential privacy. There is an exact source priority queue with private operation codes whose access pattern, grouped by external call, is $(2\eta,(1+e^\eta)\delta)$-differentially private under replacement of one complete external operation. A source history of length $t$ invokes exactly $3t$ target operations. The wrapper adds $O(1)$ local word operations per source call and $O(1)$ words to the target's private workspace at every instant. It increases record size and priority length by at most constant factors for polynomial integer domains.
\end{lemma}

\begin{proof}
Apply the wrapper of Section~\ref{sec:wrapper} to the target queue, with source operation bound $T$ and priority domain $[U]$.

The invariant is that no guard survives a round and the target's real records are exactly the source queue's records, with the indicated priority offset. A guard is smaller than every real record and every dummy, so an insertion round removes its guard. An extraction round removes a real minimum whenever one exists, because all dummies are larger than all real records. When there is no real record, it removes a dummy and reports an empty source queue. Residual dummies never affect a later real minimum. Insertion-order tie breaking is preserved because every real record occupies the first target insertion of its source round. The invariant proves correctness for arbitrary interleavings.

Now consider two equal-length source histories differing in one operation. Replacing the fields of an insertion changes only its first target insertion. Replacing an insertion by an extraction, or conversely, changes at most the two target insertions in that round. All later target inputs are fixed functions of the corresponding source inputs and the public round indices, independent of previous answers. Thus the encoded target histories differ in at most two insertion records and have identical public types. Denote these target histories by $H$ and $H'$, and write $\view(H)$ for the target core's access pattern, omitting the fixed algorithm superscript. Replacing these target records one at a time gives a two-edge neighboring path. The intermediate history is a valid target priority-queue history with the same type pattern, whether or not it encodes a source history. Applying the target privacy inequality twice gives, for every trace event $A$,
\begin{align*}
\Prb[\view(H)\in A]
&\le e^{2\eta}\Prb[\view(H')\in A]+(1+e^\eta)\delta.
\end{align*}
Let $V_i$ be the server trace of target call $i$. Since the wrapper makes no additional server accesses, source call $j$ has trace $V_{3j-2}\Vert V_{3j-1}\Vert V_{3j}$. This fixed grouping is postprocessing of the target access pattern, so the source trace with external call boundaries visible satisfies the same privacy bound. Tag checks, discarded guard answers, and returned source answers remain private.

After round $t$, the target contains exactly $t$ records, and its peak occupancy is at most $T+1$. Its operation bound is $3T$, and every priority is at most $U+3T+1$, still polynomial in $T$. Tags occupy constant space. The wrapper retains only the round index, the current operation code and constant-word input, and a constant-word answer. If the target uses at most $M(3T)$ private words at every instant, the source uses at most $M(3T)+O(1)$, including within a source call. Computing the target records, checking the returned tag, and translating the answer take $O(1)$ local word operations per source call. Server space, probes, and total work are consequently preserved up to constant factors and this $O(t)$ local overhead.
\end{proof}

\subsection{Work and Space Bounds}
\label{sec:probe-accounting}

The two main costs are segment construction and prefetching. Every core operation belongs to at most one created segment per level. Each such segment takes expected linear work in its public capacity. Prefetching reads each bin at most once during that segment's lifetime. We show that local set updates, metadata, and initialization fit in the same bound.

\begin{lemma}[Resource bounds]
\label{lem:resources}
Fix a constant $\eta>0$ and $0<\delta<1$ with $\log(1/\delta)=O(\ell^2)$. For every fixed core history of actual length $t\le T$, the expected total work is $O(t(1+\log T/\log\log T))$, and the total probe count on every execution has the same bound. The core uses $O(\log^{14}T)$ private words at every instant, including temporary workspace, and $O(T(1+\log T/\log\log T))$ server words. All actual initialization is included.
\end{lemma}

\begin{table}[!hb]
\makeatletter
\long\def\@makecaption#1#2{%
  \vskip\abovecaptionskip
  \sbox\@tempboxa{#1. #2}%
  \ifdim\wd\@tempboxa>\hsize
    #1. #2\par
  \else
    \global\@minipagefalse
    \hb@xt@\hsize{\hfil\box\@tempboxa\hfil}%
  \fi
  \vskip\belowcaptionskip}
\makeatother
\centering
\begin{tabularx}{\textwidth}{@{}Xl@{}}
\toprule
{\bf Private region} & {\bf Maximum words} \\
\midrule
Segment local sets $\mathcal{C}_r$ and their linked-list representations & $O(KB)$ \\
Global integer index for $\bigcup_r\mathcal{C}_r$ and live stash records & $O(KB)$ \\
Current epoch's original records and live bits & $O(B)$ \\
One boundary-bin remainder and cursor per merge source & $O(KB)$ \\
Current input and output bins and temporary sorting set & $O(B)$ \\
Compressed-trie stack and checkpoint tuple buffer & $O(h+R_{\mathrm{slot}})$ \\
Segment descriptors, scan counters, search stacks, commit state & $O(K+\log V)$ \\
Integer-set constants and initialization state & $O(\log w)$ \\
Noise sampler, current input and answer & $O(1)$ \\
\bottomrule
\end{tabularx}
\caption{Every private region is charged, including temporary work. Long directories, summaries, sparse event arrays, the implicit segment tree, and historical segments reside on the server. Here $B=O(\ell^{12})$, $K=O(\ell^2)$, and $R_{\mathrm{slot}}=h(1+\lfloor 2B/(H+1)\rfloor)$.}
\label{tab:space}
\end{table}

\begin{proof}
\noindent\textit{Total work and probes.}
The leading cost is linear segment construction at each level. For an actual history of length $t\le T$, each core operation belongs to at most one created segment per level, including its temporary level-$0$ segment, whether or not that operation inserts a record. The total created capacity is therefore $O(t(d+1))$. With growth factor $\ell$, the number of levels is $d+1=O(1+\ell/\log\ell)$. The remaining charges below show that boundary searches, metadata updates, and local set maintenance fit within the same bound.

For a segment of capacity $m_0$, Lemmas~\ref{lem:merge} and~\ref{lem:streaming-summary} give $O(m_0)$ probes on every execution and $O(m_0)$ expected local word operations for construction and summary generation. This charge includes padded scans, output and directory initialization, fixed checkpoint writes, and server-side event sorting. Only the work spent on capped sampling is analyzed in expectation. The parameter bounds below give $KB=w^{O(1)}$, so Lemma~\ref{lem:small-integer-sets} makes the local index and bin operations constant per record.

During a segment's lifetime, every physical bin is newly prefetched at most once because its frontier never moves backward. Each record added to $\mathcal{C}_r$ is inserted once into $\mathcal{I}$ and the linked list representing $\mathcal{C}_r$. Extraction deletes it from both structures. At retirement, $\mathcal{C}_r$ identifies the surviving entries to delete from $\mathcal{I}$. Each deletion is charged to its earlier insertion, and each insertion is charged to the newly prefetched physical slot. Stash insertions contribute only $O(t)$ additional index operations. Lemma~\ref{lem:small-integer-sets} makes their total local work $O(t(d+1))$. Dead original fields remain available in the stash until its epoch closes, without retaining their entries in $\mathcal{I}$. In particular, an epoch does not rescan unused private storage.

The total number of weighted summary entries $(z_{r,i},\omega_{r,i})$ created or removed is $O(t(d+1)/s)$, absorbing the one residual entry per segment because every created segment has capacity at least $B\ge s$. Each causes $O(\log V)$ metadata probes and word operations. Computing all bin boundaries contributes $O(t(d+1)K\log^2T/B)$. At each of at most $t/B$ epoch boundaries, cutoff search costs $O(\log V)$ and directory searches across the current segments cost $O(K\log T)$. Public extraction-root updates cost $O(t)$. The lazy segment tree and all these searches use constant local word work per visited node or probe. Consequently, both the expected total work and the pathwise total probe count are bounded by
\begin{align}
O\!\left(t(d+1)+t+
\frac{t(d+1)K\log^2T}{B}+
\frac{t(d+1)\log V}{s}+
\frac{tK\log T}{B}\right)
=O(t(d+1)).
\label{eq:total-probes}
\end{align}

\medskip
\noindent\textit{Parameter bounds.}
For fixed $\eta$ and $\log(1/\delta)=O(\ell^2)$, the definitions give
\begin{align*}
d+1&=\Theta(1+\ell/\log\ell),&
K&=O(\ell^2),&
L&=O(\ell^2),&
R_0&=O(\ell^2),\notag\\
H&=O(\ell^4),&
Z&=O(\ell^6),&
KZ&=O(\ell^8),\notag\\
B&=O(\ell^{12}),&
E&=O(B),&
R_{\mathrm{slot}}&=O(B).
\end{align*}
Here $h=O(\ell)$, $L=O(h(d+1))$, and $H=LR_0$. The checkpoint bound follows from $R_{\mathrm{slot}}\le h+\frac{2Bh}{H+1}$ and $H=LR_0$ with $L\ge4c_\eta h(d+1)$ and $R_0\ge\ell^2$. Thus $KB=O(\ell^{14})=w^{O(1)}$, which justifies the constant-time local integer-set operations.

The choices $s=\ell^6$ and $B\ge\ell^{12}$ absorb the search terms in~\eqref{eq:total-probes}. The condition $B\ge4KZ+4$ guarantees the bin and local set bounds.

\medskip
\noindent\textit{Initialization and arithmetic.}
The algorithm uses word multiplication, comparisons, shifts, and Boolean operations. It also uses most-significant-bit queries and independent uniform random words.

Under the initialization rules of Section~\ref{sec:operations}, private-heap operations cost $O(\log B)=O(d+1)$ each. The first completed epoch pays for the $O(\log w)$ integer-set setup and $O(B)$ transfer. Public parameter computation costs $O(d+\log w)=O(d+1)$ word operations on the first invocation. Record slots and auxiliary structures incur initialization costs only when used, and all masks and supporting structures are included in private space.

The division and parameter-arithmetic implementations are given in Appendix~\ref{app:word-ram}. Each segment-specific division costs $O(w)$ operations and is charged to a created segment of capacity at least $B=\Omega(\ell^{12})$. The quantization sweep advances its next threshold by addition, so it performs no division per event.

All historical data segments together occupy $O(T(d+1))$ server words. Their directories, summaries, fixed sparse arrays, and endpoint-event workspaces fit within the same total segment capacity bound, even if old versions are retained. The number of lazily allocated nodes of the implicit segment tree is at most their total update and search probes, also $O(T(d+1))$ by~\eqref{eq:total-probes}. Every actual initialization write is already charged. Server addresses therefore remain polynomial in $T$, as required by the word size. The server is passive, so its work is proportional to the charged reads and writes. 

\medskip
\noindent\textit{Private space at every instant.}
Table~\ref{tab:space} accounts for all private memory, including the local integer sets and their supporting structures. Registers contain quantities of polynomial magnitude in $T$, so each occupies a constant number of words. The integer sets have linear space by Lemma~\ref{lem:small-integer-sets}. The descriptor and list links of each record in $\mathcal{C}_r$ occupy constant space, and a removed record is unlinked immediately. Also, $R_{\mathrm{slot}}\le h+2Bh/(H+1)=O(B)$ because $H=LR_0$, $L\ge4c_\eta h(d+1)$, and $R_0\ge\ell^2$. Even when the old queue state, new parent buffers, and temporary sorting set coexist, their combined size is $O(KB)=O(\ell^{14})$.

The space bound holds for every outcome of the bounded samples. The additive privacy parameter arises from the noise distribution, while correctness and space are guaranteed on every execution.

\end{proof}

\begin{proof}[Proof of Theorem~\ref{thm:main}]
Let the external operation bound be $T$, and write $\ell_T=\lceil\log T\rceil$. Instantiate the core with operation bound $T'=3T$ and parameters
\begin{align*}
\eta=\eps/2,\qquad
\kappa\ge1+e^\eta\ \text{a fixed integer},\qquad
\delta=\delta_T/\kappa.
\end{align*}
Choose a fixed positive integer $c_\eta\ge1/\eta$ and write $\ell'=\lceil\log T'\rceil$. Since $\log(1/\delta)=\ell_T^2+\log\kappa=O((\ell')^2)$, Lemmas~\ref{lem:correctness}, \ref{lem:core-privacy}, and~\ref{lem:resources} apply with core operation bound $T'$. Lemma~\ref{lem:comparison-hide-types} gives $(2\eta,(1+e^\eta)\delta)$ privacy, with $2\eta=\eps$ and $(1+e^\eta)\delta\le\kappa\delta=\delta_T$.
An actual external history of length $t$ produces exactly $3t$ core calls. Since $\ell'=\Theta(\ell_T)$, the work, probe, and space bounds change by only constant factors. The wrapper uses $O(1)$ additional private words and local work per external call and preserves exactness and termination.
\end{proof}

\subsection{Dependence on Privacy Parameters}
\label{sec:parameterized-analysis}
We extend Theorem~\ref{thm:main} to privacy parameters that may depend on $T$. The private integer sets can then grow beyond $w^{O(1)}$ keys, so we use the general bound on the cost of each integer-set operation.

\begin{theorem}[Explicit dependence on the privacy parameters]
\label{thm:parameterized}
Fix a constant $c_{\mathrm{priv}}\ge1$ and a polynomial bound on the integer priority universe. Let $T>2$, $\eps>0$, and $0<\delta<1$, and define
\begin{align*}
\ell&=\lceil\log T\rceil, &
\bar{\eps}&=\min\{\eps,1\}, &
R&=\ell^2+\log (1/\delta),\notag\\
M_*&=\ell^{14}+\frac{\ell^7R}{\bar{\eps}}, &
g_*&=1+\frac{\log (2+M_*)}{\log \ell}.
\end{align*}
Suppose $R/\bar{\eps}\le T^{c_{\mathrm{priv}}}$. The privacy parameters are represented by the public integers $c_\eta=\lceil2/\bar{\eps}\rceil$ and $\nu=\lceil\log (3/\delta)\rceil$. These suffice to compute the construction parameters using integer arithmetic.

There is an exact online $(\eps,\delta)$-differentially oblivious priority queue under replacement of one complete operation. For every fixed history of actual length $t\le T$, its expected total work is
\begin{align*}
O\!\left(t\left(1+\frac{\ell}{\log\ell}\right)g_*\right),
\end{align*}
and its total probe count on every execution is $O(t(1+\ell/\log\ell))$. It uses $O(M_*)$ private words at every instant, including temporary workspace, and $O(T(1+\ell/\log\ell))$ server words. Every execution terminates with exact answers, and all actual initialization is charged. The word length is $w=\Theta(\log T)$. Its constant factor and the constants in these bounds may depend on $c_{\mathrm{priv}}$ and the fixed priority-universe exponent, but not on $\eps$ or $\delta$.
\end{theorem}

For fixed $\eps>0$ and $\log(1/\delta)=O(\ell^2)$, we have $M_*=O(\ell^{14})$ and $g_*=O(1)$, recovering the asymptotic work bound of Theorem~\ref{thm:main}. Appendix~\ref{app:parameterized} uses the same privacy and access-pattern analysis, together with the general bound on the cost of integer-set operations, to derive the stated work and space bounds as functions of $\eps$ and $\delta$.

\section{Lower Bounds}
\label{sec:lower-bounds}

We prove a matching probe lower bound in the balls-and-bins model of Section~\ref{sec:balls-and-bins}, with observable call boundaries. The bound holds even when local computation on priorities and metadata is arbitrary and free.

Formally, we have the following theorem.

\begin{theorem}[Lower bound]
\label{thm:lb-main}
Let $2\le n\le T/2$ and $1\le M\le n/2$. Consider an online priority queue $\mathcal{Q}$ in the balls-and-bins model with at most $M$ private words of $w=\Theta(\log T)$ bits each, and a priority domain containing $[n]$. The queue terminates with exact answers on every execution. Payloads are one-word tokens supplied only when their insertions are invoked and returned only from private memory.

For any privacy parameters $\eps=O(1)$ and $\delta=o(1)$, if $\mathcal{Q}$ is $(\eps,\delta)$-DO under replacement of the record supplied by one insertion, then there exists a fixed sequence of $O(n)$ operations on which $\mathcal{Q}$ incurs $\Omega\!\left(\frac{n\log n}{\log (M+2)}\right)$ expected total probes.
\end{theorem}

We prove this theorem in the following subsections. Intuitively, our proof compares a lower and an upper bound on the conditional entropy of the output permutation given the trace. Differential privacy leaves substantial uncertainty about that permutation. With opaque payloads, a fixed trace and limited private storage can realize only a bounded number of permutations.

\subsection{Residual Permutation Entropy}
\label{sec:lb-entropy}

Fix distinct tokens $b_1,\ldots,b_n$. Let $X_1,\ldots,X_n$ be independent and uniform on $[n]$, and consider the random operation sequence $\sigma$ consisting of $2n$ operations
\begin{align}
\sigma=\bigl(\Insert(X_1,b_1),\ldots,\Insert(X_n,b_n),
\underbrace{\Extract(),\ldots,\Extract()}_{n\text{ calls}}\bigr).
\label{eq:lb-history}
\end{align}
Let $X=(X_1,\ldots,X_n)$, and let $P$ be the total number of probes made by $\mathcal{Q}$ on $\sigma$. The sequence of returned payloads is $(b_{\Pi(1)},\ldots,b_{\Pi(n)})$, where the permutation $\Pi$ is determined by $X$ and insertion-order tie breaking. Probabilities and expectations in the proof range over both $X$ and the implementation's randomness unless stated otherwise. Entropy and mutual information are measured in bits.

Letting 
\begin{align*}
  \gamma_n=\frac{\eps+1}{\ln2}+\frac{\delta\log n}{1-e^{-1}},
\end{align*}
the following lemma shows that $\gamma_n$ upper bounds the information that the trace reveals about one priority after the preceding priorities have been fixed. 

\begin{lemma}[Residual permutation entropy]
\label{lem:lb-entropy}
Under the privacy hypothesis of Theorem~\ref{thm:lb-main}, the output permutation $\Pi$ and access pattern on $\sigma$ satisfy
\begin{align}
\mathrm{H}(\Pi\mid \view^\mathcal{Q}(\sigma))\ge n(\log n-2-\gamma_n).
\label{eq:lb-entropy}
\end{align}
This entropy bound remains valid without payload indivisibility.
\end{lemma}

\begin{proof}
Let $Y$ be the nondecreasing rearrangement of $X$. The pair $(Y,\Pi)$ determines $X$, since $X_{\Pi(j)}=Y_j$. There are $\binom{2n-1}{n}<2^{2n}$ possible multisets $Y$. Consequently,
\begin{align}
\mathrm{H}(\Pi)
&\ge \mathrm{H}(X)-\mathrm{H}(Y)
\ge n\log n-2n.
\label{eq:lb-permutation-entropy}
\end{align}

To bound the information in the trace, fix $i$ and a value $u$ of $X_{<i}=(X_1,\ldots,X_{i-1})$. For $x\in[n]$, let $\mu_x$ be the conditional law of $\view^\mathcal{Q}(\sigma)$ given $X_{<i}=u$ and $X_i=x$, averaging over the independent suffix $X_{>i}$. Let $\mu=n^{-1}\sum_{x=1}^n\mu_x$. For any fixed suffix, changing $x$ changes only one insertion. Averaging the privacy inequality over that suffix and then over the alternative value of $x$ gives, for every trace event $A$,
\begin{align}
\mu_x(A)\le e^\eps\mu(A)+\delta.
\label{eq:lb-mixture}
\end{align}
Also, $\mu(v)\ge\mu_x(v)/n$ for every trace $v$.

Let $B_x=\{v:\mu_x(v)>e^{\eps+1}\mu(v)\}$. Applying~\eqref{eq:lb-mixture} to $B_x$ yields
\begin{align*}
\mu_x(B_x)
\le e^\eps\mu(B_x)+\delta
\le e^{-1}\mu_x(B_x)+\delta
\quad\text{and}\quad
\mu_x(B_x)\le\frac{\delta}{1-e^{-1}}.
\end{align*}
Outside $B_x$, the log likelihood ratio is at most $(\eps+1)/\ln2$. On $B_x$, it is at most $\log n$. Thus the relative entropy, with base-two logarithms, satisfies
\begin{align}
D_{\mathrm{KL}}(\mu_x\Vert\mu)
&=\sum_{v:\mu_x(v)>0}\mu_x(v)\log\frac{\mu_x(v)}{\mu(v)}
\notag\\
&\le\frac{\eps+1}{\ln2}+\mu_x(B_x)\log n
\le\gamma_n.
\label{eq:lb-kl}
\end{align}
All traces are finite by termination, so their range is countable. The mixture domination above bounds the likelihood ratio even when the probe count is unbounded.

Since $X_i$ remains uniform conditional on $X_{<i}=u$, averaging~\eqref{eq:lb-kl} over $x$ bounds $\mathrm{I}(X_i;\view^\mathcal{Q}(\sigma)\mid X_{<i}=u)$ by $\gamma_n$. Averaging over $u$ and using the chain rule gives
\begin{align*}
\mathrm{I}(X;\view^\mathcal{Q}(\sigma))
=\sum_{i=1}^n\mathrm{I}(X_i;\view^\mathcal{Q}(\sigma)\mid X_{<i})
\le n\gamma_n.
\end{align*}
The permutation $\Pi$ is a function of $X$, so data processing bounds $\mathrm{I}(\Pi;\view^\mathcal{Q}(\sigma))$ by the same quantity. Subtracting from~\eqref{eq:lb-permutation-entropy} proves~\eqref{eq:lb-entropy}.
\end{proof}

\subsection{Routing Under a Fixed Trace}
\label{sec:lb-routing}

The entropy bound uses only privacy and exactness. The next lemma uses the opaque-token restriction and the private-space bound to bound the number of output permutations supported by a fixed trace.

\begin{lemma}[Fixed-trace routing]
\label{lem:lb-routing}
For the input family in~\eqref{eq:lb-history}, a fixed server trace $v$ with $q$ probes supports at most
\begin{align}
[2(M+2)]^{q+2n}
\label{eq:lb-routing-count}
\end{align}
output permutations in the opaque-payload model. Hence
\begin{align}
\mathrm{H}(\Pi\mid\view^\mathcal{Q}(\sigma))
\le(\E[P]+2n)\log(2(M+2)).
\label{eq:lb-routing-entropy}
\end{align}
\end{lemma}

\begin{proof}
The trace specifies each probe's address, read/write type, and call. For counting, we ignore the observed call boundaries and retain only the ordered probes. The $n$ input arrivals and $n$ outputs have a fixed relative order, with tokens arriving as $b_1,\ldots,b_n$ before the outputs. The number of interleavings of these $2n$ events with the $q$ probes is at most
\begin{align}
\binom{q+2n}{2n}\le2^{q+2n}.
\label{eq:lb-interleavings}
\end{align}
Some interleavings may violate the observed call boundaries or the online interface. Counting all of them only enlarges the bound.

Fix an interleaving. We describe each execution by one symbol from an alphabet of size $M+2$ per event. The description determines an auxiliary set $S$ of at most $M$ token identities and a token label for each server word. A label is either one of $b_1,\ldots,b_n$ or the symbol $*$ for a word containing none of these tokens. Initially $S$ is empty and every server label is $*$. The invariant is that the decoded server labels agree with the execution and that $S$ contains every input token currently held by the client. The decoder orders the elements of $S$ by their input indices, so a choice of a resident token has an unambiguous encoding.

Only an input arrival or a read can supply a new input token to the client. The token supplied by an arrival is fixed by the input order. The token supplied by a read is fixed by the current label at the probed address. In either case, the decoder forms $S^+$ by adjoining that token to $S$, if the label is not $*$. Since $|S^+|\le M+1$, at most one deletion is needed to restore the bound $M$. If $|S^+|=M+1$, at least one member is absent from the actual client state after the event, because that state holds at most $M$ distinct tokens. The description identifies one such member to delete by its position in $S^+$ ordered by input index. If $|S^+|\le M$, it specifies no deletion. There are at most $M+2$ choices, and the invariant is preserved. Tokens that the actual computation discards may remain in $S$.

For a write, the description selects an element of $S$ or $*$ as the new label at the fixed server address. There are at most $M+1$ choices. For an output, it selects an element of $S$, which determines the next token of the output permutation. These events need not remove anything from $S$. Between events, private computation can copy or discard tokens but cannot introduce a new identity. It therefore preserves containment in $S$ without further symbols. This includes arbitrary computation on priorities and metadata, whose storage we ignore in the auxiliary description.

Induction on the event sequence shows that every legal execution has such a description. Conversely, an interleaving and a description determine all auxiliary sets, server labels, and output identities from the initial state. Thus, even when private random coins and metadata differ between executions, each description determines at most one output permutation. The number of descriptions is at most $(M+2)^{q+2n}$. Multiplication by~\eqref{eq:lb-interleavings} proves~\eqref{eq:lb-routing-count}.

For each fixed $v$, the entropy of $\Pi$ conditional on $\view^\mathcal{Q}(\sigma)=v$ is at most the logarithm of its support size. Averaging this bound over $\view^\mathcal{Q}(\sigma)$ proves~\eqref{eq:lb-routing-entropy}. The output identities appear only in the counting argument, not in the observed trace.
\end{proof}

\subsection{Proof of the Main Lower Bound}
\label{sec:lb-conclusion}

Finally, we can provide a full proof for Theorem~\ref{thm:lb-main}.

\begin{proof}[Proof of Theorem~\ref{thm:lb-main}]
If $\E[P]=\infty$ under~\eqref{eq:lb-history}, some fixed input in this finite family already has infinite expected cost. Otherwise, Lemmas~\ref{lem:lb-entropy} and~\ref{lem:lb-routing} apply to the same output permutation and observed trace, giving
\begin{align}
n(\log n-2-\gamma_n)
&\le(\E[P]+2n)\log(2(M+2)),\notag\\
\E[P]
&\ge\frac{n(\log n-2-\gamma_n)}{\log(2(M+2))}-2n.
\label{eq:lb-quantitative}
\end{align}

For $\eps=O(1)$ and $\delta=o(1)$, we have $\gamma_n=o(\log n)$. For all sufficiently large $n$, the numerator factor $\log n-2-\gamma_n$ is at least $\tfrac12\log n$. Set $r_n=\log n/\log(2(M+2))$. If $r_n\ge8$, equation~\eqref{eq:lb-quantitative} gives $\E[P]\ge nr_n/4$.

For $r_n<8$, a storage bound suffices. After the insertion phase, every token must have a recoverable copy, since every token will be output. At most $M$ distinct tokens can be held privately, so at least $n-M$ must be on the server. No input token is initially there, and each write places at most one token in a server word. Consequently $P\ge n-M\ge n/2$ on every execution, which also gives $\E[P]=\Omega(nr_n)$ in this range. Since $\log(2(M+2))=\Theta(\log(M+2))$, both cases yield the bound in Theorem~\ref{thm:lb-main}.

This expectation averages over the finite set of priority vectors $X\in[n]^n$. At least one fixed vector has expected probe count at least that average, with expectation now only over the implementation's randomness. Its history has the required form.
\end{proof}

\begin{corollary}[Optimal cost with opaque payloads]
\label{cor:lb-tightness}
For the privacy parameters of Theorem~\ref{thm:main} and a priority domain containing $[\lfloor T/2\rfloor]$, the construction has optimal expected amortized probe complexity and total word-RAM work among exact opaque-record implementations with polylogarithmic private space:
\begin{align*}
\Theta(\log T/\log\log T).
\end{align*}
\end{corollary}

\begin{proof}
Take $n=\lfloor T/2\rfloor$. Every fixed polylogarithmic private-space bound satisfies $M\le n/2$ for sufficiently large $T$ and $\log(M+2)=O(\log\log T)$. Complete-operation privacy implies privacy under replacement of an insertion's fields. Theorem~\ref{thm:lb-main} therefore gives $\Omega(T\log T/\log\log T)$ expected probes on a fixed history of $2n=\Theta(T)$ operations. This also lower-bounds total work.

For the upper bound, the construction of Theorem~\ref{thm:main} can store each payload intact. Its summaries and integer-set indices operate on keys, counters, and pointers. The stash, bins, merges, and local sets only retain or copy the payloads. The wrapper's tags can be stored as separate metadata, using a constant number of words per record. Thus the construction meets the opaque-payload restriction without changing its asymptotic costs, and Theorem~\ref{thm:main} supplies the matching upper bounds.
\end{proof}

The optimality claim concerns expected amortized probes and total work, not private-space or server-space optimality or worst-case time for an individual call. With arbitrary payload encodings, Lemma~\ref{lem:lb-entropy} still applies, but the fixed-trace routing bound is not established.

\section{Related Work and Discussions}
\label{sec:related}

\paragraph{Oblivious algorithms and data structures.}
ORAM gives a general method for hiding the access patterns of RAM computations~\cite{GoldreichOstrovsky1996,StefanovEtAl2013PathORAM}. Wang et al.~\cite{WangEtAl2014ODS} use pointer and locality techniques to obtain specialized oblivious data structures, including maps, stacks, and priority queues. Such constructions exploit the restricted interfaces of data structures to improve on generic simulations. Jacob, Larsen, and Nielsen~\cite{JacobLarsenNielsen2019} establish logarithmic access lower bounds for stacks, queues, deques, and priority queues in a model where the probes of each call are observed separately. Their model allows arbitrary free local computation, so the bounds concern data movement itself.

\paragraph{Differential obliviousness.}
Chan et al.~\cite{ChanCCMS2019,ChanChungMaggsShi2022} establish privacy and complexity tradeoffs for sorting, merging, and range queries. Beimel, Nissim, and Zaheri~\cite{BeimelNissimZaheri2019} show further separations from full obliviousness for tasks including graph property testing. Komargodski and Shi~\cite{KomargodskiShi2021} give differentially oblivious Turing-machine simulations and online stacks and queues with $O(\log\log T)$ amortized overhead in the Turing-machine model. In contrast, Persiano and Yeo~\cite{PersianoYeo2019,PersianoYeo2023} prove logarithmic lower bounds for differentially private RAMs and develop lower-bound frameworks for other data structures. These results show that the benefits of DO depend on both the functionality and the computational model.

Composition of DO computations requires care because the outputs of a first computation on neighboring inputs need not remain neighboring. Zhou et al.~\cite{ZhouShiChanMaimon2023} address this issue through neighbor-preserving differential obliviousness, and subsequent work establishes advanced composition theorems~\cite{ZhouZhaoChanShi2024}. Our analysis composes DP summaries of fixed insertion histories and obtains the server trace by postprocessing these releases.

\paragraph{Priority queues.}
Thorup~\cite{Thorup2000} studies integer priority queues with sublogarithmic operation bounds on a word-RAM. His later reduction~\cite{Thorup2007} relates priority queues to sorting, while Wei and Yi~\cite{WeiYi2014} establish an analogous connection in external memory. Beyond the logarithmic oblivious constructions discussed above, Thie{\ss}en and Vahrenhold~\cite{ThiessenVahrenhold2024} obtain an oblivious priority queue with perfect security and correctness, $O(\log T)$ amortized runtime, and $O(1)$ private memory cells. Maheshwari and Shi~\cite{MaheshwariShi2026} study external-memory oblivious priority queues with decrease-key operations and optimize both I/O cost and total work. Our interface consists of insertion and minimum extraction.

\paragraph{Private binning and merging.}
Private summaries and padded bins already appear in the merging algorithms of Chan et al.~\cite{ChanChungMaggsShi2022} and SWAT~\cite{ZhengEtAl2024SWAT}. SWAT discusses direct merging of several binned inputs, while its analyzed construction uses iterative pairwise merging. Our construction combines summary-guided merging with summaries of original insertions and reflected live-prefix counts. This combination supports exact online extraction and bounds both total work and private workspace.

\section{Conclusion}
\label{sec:conclusion}

We have established tight bounds for exact online differentially oblivious integer priority queues in the balls-and-bins model.
For polynomially bounded priorities, constant $\eps>0$, and $\delta=\exp(-\Theta(\log^2 T))$, our construction uses polylogarithmic private memory and incurs $O(\log T/\log\log T)$ expected amortized total work per operation.
The same amortized bandwidth bound holds on every execution.
Differentially private summaries of insertion histories and reflection at zero enable exact minimum extraction.
A matching probe lower bound, already for priorities in $[T]$, combines residual permutation entropy with constraints on routing opaque records through a fixed trace.

It remains open whether the lower bound extends to implementations that jointly encode several payloads.
Further questions are whether the private space can be reduced while preserving the optimal operation cost, and whether comparable worst-case bounds can be obtained for individual operations.

\section*{AI disclosure}
The main results of this work were developed primarily by the authors. We use GPT-6 Astra to construct part of the lower bound results, validate the results, and draft part of the proofs. Besides, GPT was also used for polishing the texts. The authors take full responsibility for the content of the manuscript.

\bibliographystyle{unsrt}
\bibliography{ref}

@inproceedings{ChanCCMS2019,
  author    = {Chan, T.-H. Hubert and Chung, K.-M. and Maggs, B. M. and Shi, E.},
  title     = {Foundations of Differentially Oblivious Algorithms},
  booktitle = {Proceedings of the 30th Annual ACM--SIAM Symposium on Discrete Algorithms (SODA)},
  pages     = {2448--2467},
  year      = {2019}
}

@inproceedings{Dwork06D,
  author    = {Dwork, Cynthia},
  title     = {Differential privacy},
  booktitle = {Proceedings of ICALP},
  year      = {2006}
}

@inproceedings{PersianoYeo2023,
  author    = {Persiano, G. and Yeo, K.},
  title     = {Lower Bound Framework for Differentially Private and Oblivious Data Structures},
  booktitle = {Advances in Cryptology---EUROCRYPT 2023},
  pages     = {487--517},
  year      = {2023}
}

@inproceedings{BeimelNissimZaheri2019,
  author    = {Beimel, A. and Nissim, K. and Zaheri, M.},
  title     = {Exploring Differential Obliviousness},
  booktitle = {APPROX/RANDOM 2019},
  series    = {LIPIcs},
  volume    = {145},
  pages     = {65:1--65:20},
  year      = {2019}
}

@inproceedings{PersianoYeo2019,
  author    = {Persiano, G. and Yeo, K.},
  title     = {Lower Bounds for Differentially Private {RAM}s},
  booktitle = {Advances in Cryptology---EUROCRYPT 2019},
  series    = {LNCS},
  volume    = {11476},
  pages     = {404--434},
  year      = {2019}
}

@inproceedings{KomargodskiShi2021,
  author    = {Komargodski, I. and Shi, E.},
  title     = {Differentially Oblivious {Turing} Machines},
  booktitle = {ITCS 2021},
  series    = {LIPIcs},
  volume    = {185},
  pages     = {68:1--68:19},
  year      = {2021}
}

@inproceedings{ZhouShiChanMaimon2023,
  author    = {Zhou, M. and Shi, E. and Chan, T.-H. H. and Maimon, S.},
  title     = {A Theory of Composition for Differential Obliviousness},
  booktitle = {Advances in Cryptology---EUROCRYPT 2023},
  series    = {LNCS},
  volume    = {14006},
  pages     = {3--34},
  year      = {2023}
}

@inproceedings{ZhouZhaoChanShi2024,
  author    = {Zhou, M. and Zhao, M. and Chan, T.-H. H. and Shi, E.},
  title     = {Advanced Composition Theorems for Differential Obliviousness},
  booktitle = {ITCS 2024},
  series    = {LIPIcs},
  volume    = {287},
  pages     = {103:1--103:24},
  year      = {2024}
}

@article{ChanChungMaggsShi2022,
  author    = {Chan, T.-H. Hubert and Chung, Kai-Min and Maggs, Bruce M. and Shi, Elaine},
  title     = {Foundations of Differentially Oblivious Algorithms},
  journal   = {Journal of the ACM},
  volume    = {69},
  number    = {4},
  pages     = {27:1--27:49},
  year      = {2022}
}

@inproceedings{JacobLarsenNielsen2019,
  author    = {Jacob, Riko and Larsen, Kasper Green and Nielsen, Jesper Buus},
  title     = {Lower Bounds for Oblivious Data Structures},
  booktitle = {Proceedings of the Thirtieth Annual ACM-SIAM Symposium on Discrete Algorithms},
  pages     = {2439--2447},
  year      = {2019}
}

@inproceedings{JafargholiLarsenSimkin2021,
  author    = {Jafargholi, Zahra and Larsen, Kasper Green and Simkin, Mark},
  title     = {Optimal Oblivious Priority Queues},
  booktitle = {Proceedings of the 2021 ACM-SIAM Symposium on Discrete Algorithms},
  pages     = {2366--2383},
  year      = {2021}
}

@inproceedings{Shi2020PathObliviousHeap,
  author    = {Shi, Elaine},
  title     = {Path Oblivious Heap: Optimal and Practical Oblivious Priority Queue},
  booktitle = {2020 IEEE Symposium on Security and Privacy},
  pages     = {842--858},
  year      = {2020}
}

@article{Thorup2007,
  author    = {Thorup, Mikkel},
  title     = {Equivalence between Priority Queues and Sorting},
  journal   = {Journal of the ACM},
  volume    = {54},
  number    = {6},
  pages     = {28},
  year      = {2007}
}

@inproceedings{WeiYi2014,
  author    = {Wei, Zhewei and Yi, Ke},
  title     = {Equivalence between Priority Queues and Sorting in External Memory},
  booktitle = {Algorithms - ESA 2014},
  series    = {Lecture Notes in Computer Science},
  volume    = {8737},
  pages     = {830--841},
  year      = {2014}
}

@inproceedings{PatrascuThorup2014,
  author    = {P{\u a}tra{\c s}cu, Mihai and Thorup, Mikkel},
  title     = {Dynamic Integer Sets with Optimal Rank, Select, and Predecessor Search},
  booktitle = {2014 IEEE 55th Annual Symposium on Foundations of Computer Science},
  pages     = {166--175},
  year      = {2014}
}

@article{ZhengEtAl2024SWAT,
  author    = {Zheng, Leqian and Xu, Lei and Wang, Cong and Wang, Sheng and Hu, Yuke and Qin, Zhan and Li, Feifei and Ren, Kui},
  title     = {{SWAT}: A System-Wide Approach to Tunable Leakage Mitigation in Encrypted Data Stores},
  journal   = {Proceedings of the VLDB Endowment},
  volume    = {17},
  number    = {10},
  pages     = {2445--2458},
  year      = {2024}
}

@inproceedings{MaheshwariShi2026,
  author    = {Maheshwari, Arya and Shi, Elaine},
  title     = {Oblivious Priority Queue and Single-Source Shortest Path in the External Memory Setting},
  booktitle = {Advances in Cryptology---CRYPTO 2026},
  series    = {Lecture Notes in Computer Science},
  volume    = {16807},
  pages     = {329--360},
  year      = {2026}
}

@article{GoldreichOstrovsky1996,
  author    = {Goldreich, Oded and Ostrovsky, Rafail},
  title     = {Software Protection and Simulation on Oblivious {RAM}s},
  journal   = {Journal of the ACM},
  volume    = {43},
  number    = {3},
  pages     = {431--473},
  year      = {1996}
}

@inproceedings{StefanovEtAl2013PathORAM,
  author    = {Stefanov, Emil and van Dijk, Marten and Shi, Elaine and Fletcher, Christopher and Ren, Ling and Yu, Xiangyao and Devadas, Srinivas},
  title     = {Path {ORAM}: An Extremely Simple Oblivious {RAM} Protocol},
  booktitle = {Proceedings of the 2013 ACM SIGSAC Conference on Computer \& Communications Security},
  pages     = {299--310},
  year      = {2013}
}

@article{FredmanWillard1993,
  author    = {Fredman, Michael L. and Willard, Dan E.},
  title     = {Surpassing the Information Theoretic Bound with Fusion Trees},
  journal   = {Journal of Computer and System Sciences},
  volume    = {47},
  number    = {3},
  pages     = {424--436},
  year      = {1993}
}

@article{Williams1964,
  author    = {Williams, J. W. J.},
  title     = {Algorithm 232: Heapsort},
  journal   = {Communications of the ACM},
  volume    = {7},
  number    = {6},
  pages     = {347--348},
  year      = {1964}
}

@article{FredmanTarjan1987,
  author    = {Fredman, Michael L. and Tarjan, Robert Endre},
  title     = {Fibonacci Heaps and Their Uses in Improved Network Optimization Algorithms},
  journal   = {Journal of the ACM},
  volume    = {34},
  number    = {3},
  pages     = {596--615},
  year      = {1987}
}

@article{Thorup2000,
  author    = {Thorup, Mikkel},
  title     = {On {RAM} Priority Queues},
  journal   = {SIAM Journal on Computing},
  volume    = {30},
  number    = {1},
  pages     = {86--109},
  year      = {2000}
}

@inproceedings{WangEtAl2014ODS,
  author    = {Wang, Xiao Shaun and Nayak, Kartik and Liu, Chang and Chan, T.-H. Hubert and Shi, Elaine and Stefanov, Emil and Huang, Yan},
  title     = {Oblivious Data Structures},
  booktitle = {Proceedings of the 2014 ACM SIGSAC Conference on Computer and Communications Security},
  pages     = {215--226},
  year      = {2014}
}

@inproceedings{ThiessenVahrenhold2024,
  author    = {Thie{\ss}en, Thore and Vahrenhold, Jan},
  title     = {Optimal Offline {ORAM} with Perfect Security via Simple Oblivious Priority Queues},
  booktitle = {35th International Symposium on Algorithms and Computation (ISAAC 2024)},
  series    = {Leibniz International Proceedings in Informatics (LIPIcs)},
  volume    = {322},
  pages     = {55:1--55:18},
  year      = {2024}
}

@article{DworkRoth2014,
  author    = {Cynthia Dwork and Aaron Roth},
  title     = {The Algorithmic Foundations of Differential Privacy},
  journal   = {Foundations and Trends{\textregistered} in Theoretical Computer Science},
  volume    = {9},
  number    = {3--4},
  pages     = {211--407},
  year      = {2014}
}

\clearpage
\appendix

\section{Heavy-node enumeration}
\label{app:enumeration}

This appendix gives the iterator used in Lemma~\ref{lem:heavy-iterator} for a segment $r$. It processes original records with weight one and checkpoint points with weight zero. Heavy-node counts are private, and the iterator makes no server writes.

\paragraph{A compressed-trie iterator.} Set $v=\log V=h-1$ and normalize a record key $x$ to $y=x-1\in\{0,\ldots,V-1\}$. A dyadic node of height $k$ has interval length $2^k$. Leaves have height zero, and the root has height $v$. Read the original records from the sorted physical bins in order, treating each as a point of weight one. We will also insert zero-weight checkpoint points. Maintain a current descriptor $(z,k,n)$ consisting of a representative coordinate, a subtree height, and its exact total weight. The dyadic node represented by this descriptor has not yet been emitted. A stack frame $(z,p,n_{\rm left})$ represents an open branching node of height $p$ whose completed left part has weight $n_{\rm left}$. The frame heights strictly decrease from the bottom of the stack to its top.

To \emph{close the edge} from a current descriptor $(z,k,n)$ to an excluded ancestor height $p>k$, do nothing if $n\le H$. Otherwise emit, for each height $i=k,\ldots,p-1$, the interval containing $z$ of length $2^i$, with count $n$. These intervals are computed directly by shifts and masking. Every implicit node on this edge has the same count, so a light edge is skipped in constant time, whereas each iteration on a heavy edge emits a distinct heavy node.

The first input coordinate creates a current leaf with its given weight. An equal next coordinate adds its weight to that still-open leaf. For a new larger coordinate $y$, let $y_{\rm old}$ be the preceding coordinate and let $p=1+\lfloor\log (y_{\rm old}\mathbin{\mathrm{xor}}y)\rfloor$, computed with a most-significant-bit operation. Perform the following steps.
\begin{enumerate}[label=(\roman*),leftmargin=*]
\item While the top frame has height less than $p$, pop it, close the current edge up to its height, and replace the current descriptor by that frame's node, adding the current count to the stored left count.
\item Close the current edge up to height $p$. Push a frame of height $p$ storing its representative and count as the completed left part, and make the new coordinate $y$ the current leaf.
\end{enumerate}
After step (i), the remaining top frame, if any, has height strictly greater than $p$. Equality cannot occur: once an increasing input stream has crossed from the left child of a frame into its right child, no later coordinate can first diverge at that frame. At the end of the domain, pop all frames by the same close-and-add rule and finally close the current descriptor up to excluded height $v+1$, thereby emitting the root if it is heavy. If there is no current descriptor, there is nothing to close. Counts are combined when branches close, so each input record does not require a separate update at every ancestor.

\paragraph{Closing nodes at bin boundaries.} The checkpoint for bin $j$ must include every heavy node whose right endpoint lies in $(q_{r,j-1},q_{r,j}]$. After processing that bin, the iterator inserts a zero-weight point at the original coordinate $q_{r,j}+1$ if $q_{r,j}<V$. This sentinel closes all nodes ending at or before $q_{r,j}$, so each newly emitted node is assigned to the completed checkpoint. At the first cut equal to $V$, the iterator instead closes the entire stack and marks it finished.

Coincident points use the same open leaf: an original record at a sentinel coordinate adds its weight to that leaf, and repeated cuts add zero weight without creating branches. Bins between repeated cuts at $V$ contain no original records, and these cuts leave the finished iterator unchanged. Thus no node is closed or sampled twice.

\begin{proof}[Proof of Lemma~\ref{lem:heavy-iterator}]
The stack represents the rightmost path of the compact binary trie of the processed coordinates. The ordered scan closes a subtree only after all its records have been read, combining its child weights. Original records have weight one and sentinels have weight zero, so a subtree's completed weight is its exact record count. All implicit nodes on a compressed edge have the same count as its child. Each edge closes once and emits exactly its heavy nodes with their exact counts. Every heavy node belongs to one such edge or is the root closed at the end, giving complete enumeration without duplication.

Completing bin $i$ includes its checkpoint step. If $q_{r,i}<V$, the sentinel at $q_{r,i}+1$ closes all pending nodes whose right endpoints are at most $q_{r,i}$. The previous checkpoint has already closed those ending at or before $q_{r,i-1}$. Thus checkpoint $i$ receives exactly the heavy nodes with right endpoints in $(q_{r,i-1},q_{r,i}]$, all enumerated by completion of bin $i$. Flushing the stack at the first boundary equal to $V$ gives the same guarantee. Postorder emits nodes by increasing right endpoint, breaking ties by increasing height. Empty bins and repeated boundaries preserve these properties.

Each distinct input point creates at most one frame, and each frame is popped once. There are at most $m+m_0/B$ original records and sentinel points, giving $O(m+m_0/B)$ structural work. At each tree level, disjoint heavy intervals contain at least $H+1$ records each, so there are at most $mh/(H+1)$ heavy nodes in total. Each emitted node costs constant work, and scanning all $2m_0$ physical slots costs $O(m_0)$. This proves the claimed deterministic work bound, excluding noise sampling. The stack has at most $h-1$ frames of constant size. Together with the current descriptor and one output pair, it uses $O(h)$ private words beyond the current bin buffer. Outputs are passed individually to private computation without server writes.
\end{proof}

\section{Word-RAM implementation details}
\label{app:word-ram}

\paragraph{Primitive instructions.}
The client supports unit-time addition, subtraction, multiplication, comparisons, Boolean operations, shifts, most-significant-bit queries, and generation of an independent uniform random word. The constant in $w=\Theta(\log T)$ accommodates composite keys, server addresses, and polynomially bounded counters. All supporting constants, temporary storage, sampling state, and input and output registers are counted in private space.

\paragraph{Exact sampling with bounded work and space.} To sample $\min\{G,H\}$, start with a counter $z=0$. While $z<H$, perform a Bernoulli trial of success probability $1/L$: on success return $z$, and on failure increment $z$. If $z=H$, return $H$. Since $L$ is a power of two, each trial masks the low $\log L$ bits of a fresh uniform random word. The procedure uses constant private space, returns after at most $H$ trials on every random tape, and uses at most $L$ trials in expectation. Its distribution assigns mass $(1-q)q^j$ to $0\le j<H$ and mass $q^H$ to $H$. The sampler makes no server accesses.

\paragraph{Lazy metadata operations.}
During a descent in the implicit segment tree, the parent's lazy transformation is propagated to both children before being cleared. A missing child is initialized to its implicit zero state before receiving the transformation. The descent then uses the updated child maxima. Values and addresses are polynomially bounded in $T$ and fit in constant-sized word fields. Each suffix addition or cutoff search visits $O(\log V)$ nodes and uses constant local word work per visited node.

\paragraph{Integer arithmetic and quantization.}
For core parameter computation, $\delta$ is represented by the public integer $\nu=\lceil\log(1/\delta)\rceil$. Integer division need not be a unit-time primitive. Public quantities such as $\lfloor 2B/(H+1)\rfloor$ can be computed once when the first segment is needed, and $\lfloor m/s\rfloor$ once per segment, by $O(w)$ shift-and-subtract division. Since every created segment has $m_0\ge B=\Omega(\ell^{12})$, these costs are absorbed by its linear charge. Segment capacities and bin counts are maintained together through multiplication by the public growth factor. The quantization sweep stores the next multiple of $s$ and advances it by addition for each actual emitted position, without dividing once per event or iterating over unused coordinates.

\section{Proof of the parameterized upper bound}
\label{app:parameterized}

\begin{proof}[Proof of Theorem~\ref{thm:parameterized}]
The wrapper of Lemma~\ref{lem:comparison-hide-types} uses target operation bound $T'=3T$. A source history of actual length $t$ induces $t'=3t$ target calls. All construction parameters are evaluated at this target operation bound. Write $\ell'=\lceil\log T'\rceil$ and $d+1=\Theta(1+\ell/\log\ell)$. Take
\begin{align*}
\eta=\bar{\eps}/2,\qquad \kappa=3,\qquad
c_\eta=\lceil2/\bar{\eps}\rceil,\qquad
\nu=\lceil\log (3/\delta)\rceil.
\end{align*}
Instantiate the core with privacy parameters $(\eta,\delta/3)$, where $\delta$ is the external privacy target. Only the supplied integers $c_\eta,\nu$ are needed to compute the construction's parameters. In particular, the implementation need not evaluate an exponential or a real logarithm. Since $\eta\le1/2$, we have $1+e^\eta<3=\kappa$ and $(1+e^\eta)\delta/3\le\delta$.

We use the public parameters defined in Section~\ref{sec:layout}, now allowing $\log(1/\delta)$ to exceed $O((\ell')^2)$ and $c_\eta$ to vary with the privacy parameters. The noise-privacy proof uses the inequalities $L\ge4h(d+1)/\eta$ and $R_0\ge\nu+\lceil\log (4\kappa h(d+1))\rceil$, which still hold. Thus each affected summary is $\bigl(\eta/(d+1),\delta/(3(d+1))\bigr)$-DP, and the same composition and postprocessing proof gives $(\eta,\delta/3)$ privacy for the public-type construction. The deterministic summary error, bin-capacity, reflection, and local set arguments also continue to apply with the newly chosen parameters.

Because $\ell'=\Theta(\ell)$ and $R=\ell^2+\log (1/\delta)$, the explicit parameter dependence is
\begin{align*}
R_0&=\Theta(R), & L&=\Theta(h(d+1)/\bar{\eps}), &
H&=\Theta(h(d+1)R/\bar{\eps}),\notag\\
Z&=O\!\left(\ell^6+\frac{h^2(d+1)R}{\bar{\eps}}\right), &
B&=O\!\left(\ell^{12}+\frac{Kh^2(d+1)R}{\bar{\eps}}\right).
\end{align*}
Here $K\ell^6=O(\ell^8)$ is absorbed by the $\ell^{12}$ term. All private regions in Table~\ref{tab:space} still occupy $O(KB)$ words: in particular, $R_{\mathrm{slot}}=O(B)$ follows from the same lower bounds on $L,R_0$ and from $B\ge(\ell')^{12}$. Using $K=O(\ell^2)$, $h=O(\ell)$, and $d+1=O(\ell)$ gives
\begin{align}
KB=O\!\left(K\ell^{12}+\frac{K^2h^2(d+1)R}{\bar{\eps}}\right)
=O\!\left(\ell^{14}+\frac{\ell^7R}{\bar{\eps}}\right)=O(M_*).
\label{eq:parameterized-space}
\end{align}
The hypothesis $R/\bar{\eps}\le T^{c_{\mathrm{priv}}}$ makes all parameters, counters, event fields, and addresses polynomially bounded in $T$, with a fixed exponent. Choosing the constant in $w=\Theta(\log T)$ accordingly therefore preserves constant-word representations and word-level arithmetic throughout.

With varying privacy parameters, the local integer sets are not necessarily of size $w^{O(1)}$. Define
\begin{align*}
\rho=1+\frac{\log (2+KB)}{\log w}=O(g_*).
\end{align*}
Lemma~\ref{lem:small-integer-sets} gives $O(\rho)$ work per index operation. Sorting a level-$0$ segment or a merge output bin takes $O(B\rho)$ work and $O(B)$ space, and the lifetime accounting for local-set insertions and deletions gains the same factor. Hence a segment of capacity $m_0$ incurs $O(m_0\rho)$ local work for these tasks, while its fixed-capacity data accesses still require $O(m_0)$ probes.

Generating a summary from an already sorted segment still takes linear work in expectation. The sampling and sparse-event sorting terms are respectively
\begin{align*}
O\!\left(\frac{mhL}{H+1}\right)&=O(mh/R_0)=O(m),\\
O(P_0\log T')&=O\!\left(\frac{m_0h\log T'}{H+1}
+\frac{m_0h\log T'}{B}\right)=O(m_0).
\end{align*}
These estimates use the unchanged lower bounds $R_0\ge(\ell')^2$, $L\ge2$, and $B\ge(\ell')^{12}$. Larger noise scales do not introduce an additional sampling factor, because they reduce the number of heavy candidates. The physical scan and output bounds remain deterministic. The metadata and boundary-search terms of~\eqref{eq:total-probes} likewise remain bounded by $O(t'(d+1))=O(t(d+1))$, since $s=(\ell')^6$ and $B\ge(\ell')^{12}$ are unchanged lower bounds. The total created-segment capacity is $O(t'(d+1))=O(t(d+1))$. Summing these capacities and local set lifetimes therefore gives expected work $O(t(d+1)\rho)$, pathwise probes $O(t(d+1))$, and server space $O(T'(d+1))=O(T(d+1))$.

The initialization accounting changes only through the integer-set factor. The private heap costs $O(\log(2+B))=O(\rho\log w)=O((d+1)\rho)$ per operation, and the first completed epoch pays for $O(\log w)$ setup and an $O(B\rho)$ transfer. Given $c_\eta,\nu$, public parameter computation takes $O(d+\log w)=O(d+1)$ word operations using integer arithmetic and bit lengths. The segment-specific divisions retain their linear charges. In the private-only case $B\ge T'$, the rules of Section~\ref{sec:operations} initialize only $O(t')=O(t)$ record slots and perform no server probes. Its heap work and space therefore satisfy the claimed bounds as well. The unchanged capped samplers preserve pathwise termination.

Finally, Lemma~\ref{lem:comparison-hide-types} gives $(2\eta,(1+e^\eta)\delta/3)$ privacy, which implies $(\eps,\delta)$ privacy, and changes work, probes, and space only by constant factors. Equations~\eqref{eq:parameterized-space} and $\rho=O(g_*)$ give the claimed resource bounds.
\end{proof}

\end{document}